\documentclass[reprint,nofootinbib,amsmath,amssymb,showkeys,aps]{revtex4-1}

\usepackage{amssymb}
\usepackage{array,multirow}
\usepackage{enumerate}
\usepackage{bm}
\usepackage{amsmath}
\usepackage{graphicx}
\usepackage{xcolor}

\usepackage[flushleft]{threeparttable}

\begin{document}

\title{Coupled and uncoupled early dark energy, massive neutrinos and the cosmological tensions}

\author{Adri\`a G\'omez-Valent$^{1,2}$}\email{agvalent@roma2.infn.it}
\author{Ziyang Zheng$^{3}$}\email{zheng@thphys.uni-heidelberg.de}
\author{Luca Amendola$^{3}$}\email{l.amendola@thphys.uni-heidelberg.de}
\author{Christof Wetterich$^{3}$}\email{c.wetterich@thphys.uni-heidelberg.de}
\author{Valeria Pettorino$^4$}\email{valeria.pettorino@cea.fr}

\affiliation{$^1$ Dipartimento di Fisica, Università di Roma Tor Vergata, via della Ricerca Scientifica 1, I-00133 Roma, Italy}
\affiliation{$^2$ INFN, Sezione di Roma 2, Università di Roma Tor Vergata, via della Ricerca Scientifica 1, I-00133 Roma, Italy}
\affiliation{$^3$ Institut f\"{u}r Theoretische Physik, Ruprecht-Karls-Universit\"{a}t Heidelberg, Philosophenweg 16, D-69120 Heidelberg, Germany}
\affiliation{$^4$ AIM, CEA, CNRS, Universit{\'e} Paris-Saclay, Universit{\'e} Paris Diderot,             Sorbonne Paris Cit{\'e}, F-91191 Gif-sur-Yvette, France}

\begin{abstract}
Some cosmological models with non-negligible dark energy fractions in particular windows of the pre-recombination epoch are capable of alleviating the Hubble tension quite efficiently, while keeping the good description of the data that are used to build the cosmic inverse distance ladder. There has been an intensive discussion in the community on whether these models enhance the power of matter fluctuations, leading {\it de facto} to a worsening of the tension with the large-scale structure measurements. We address this pivotal question in the context of several early dark energy (EDE) models, considering also in some cases a coupling between dark energy and dark matter, and the effect of massive neutrinos. We fit them using the {\it Planck} 2018 likelihoods, the supernovae of Type Ia from the Pantheon compilation and data on baryon acoustic oscillations. We find that ultra-light axion-like (ULA) EDE can actually alleviate the $H_0$ tension without increasing the values of $\sigma_{12}$ with respect to those found in the $\Lambda$CDM, whereas EDE with an exponential potential does not have any impact on the tensions. A coupling in the dark sector tends to enhance the clustering of matter, and the data limit a lot the influence of massive neutrinos, since the upper bounds on the sum of their masses are too close to those obtained in the standard model. We find that in the best case, namely ULA, the Hubble tension is reduced to $\sim 2\sigma$.
\end{abstract}


\maketitle


\section{Introduction}\label{sec:intro}

The standard model of cosmology, also known as $\Lambda$CDM, relies, among other assumptions, on the existence of cold dark matter (CDM) and a constant (immutable) vacuum energy density that pervades space and is in charge of the late-time acceleration of the universe \cite{Peebles:2002gy}. Despite its apparent simplicity, this model has shown an undeniable ability to explain a wide variety of cosmological observations with an incredible accuracy, ranging from the anisotropies of the cosmic microwave background (CMB) \cite{Hinshaw:2012aka,Planck:2018vyg}, to the baryon acoustic oscillations (BAO) imprinted in the distribution of matter in the universe \cite{Eisenstein:2005su,Cole:2005sx}, and to the Hubble diagram of supernovae of Type Ia (SNIa) \cite{Perlmutter:1998np,Riess:1998cb}. This is why the $\Lambda$CDM is sometimes also referred to as the {\it concordance} model of cosmology. However, apart from very important and  long-standing theoretical problems associated to the value of the vacuum energy density (see e.g. \cite{Weinberg:1988cp} and the review \cite{SolaPeracaula:2022hpd}), we have also witnessed in the last years the stubborn persistence of significant tensions between the model and some data, which make the $\Lambda$CDM less concordant than previously thought. Two of the most prominent ones are the Hubble (or $H_0$) \cite{Verde:2019ivm} and $\sigma_8$ (or $S_8$) tensions \cite{DiValentino:2020vvd}. These are the tensions we will focus our attention on in this work, although there are others, as those linked to the CMB anomalies, see e.g. \cite{Perivolaropoulos:2021jda,Aluri:2022hzs}. The $H_0$ tension is between the Hubble constant measured by the SH0ES team model-independently with the cosmic distance ladder method \cite{Riess:2021jrx}, and the CMB-inferred value by the {\it Planck} collaboration \cite{Planck:2018vyg}, which assumes  $\Lambda$CDM. These values read, respectively, $H_0=(73.04\pm 1.04)$ km/s/Mpc and $H_0= (67.36\pm 0.54)$ km/s/Mpc\footnote{Here, as well as in the evaluation of the $S_8$ tension, we take the value obtained from the  TT,TE,EE+lowE+lensing analysis by {\it Planck} \cite{Planck:2018vyg}.}, and are in $4.8\sigma$ tension. On the other hand, the $S_8$ parameter is defined as $S_8=\sigma_8(\Omega_m^{(0)}/0.3)^{0.5}$, with $\Omega_m^{(0)}$ the current matter density fraction, $\sigma_8$ the {\it rms} of mass fluctuations at scales of $R_8=8h^{-1}$ Mpc, and $h$ the reduced Hubble parameter. The $S_8$ tension is found between, again, the $\Lambda$CDM-based inference by {\it Planck} \cite{Planck:2018vyg}, $S_8=0.832\pm 0.013$, and the value measured by some weak lensing surveys. For instance, the combined tomographic weak gravitational lensing analysis of the Kilo Degree Survey (KiDS +VIKING-450) and the Dark Energy Survey (DES-Y1) \cite{Joudaki:2019pmv} led to the measurement $S_8=0.762^{+0.025}_{-0.024}$, under the assumption of the standard model. The latter is in $2.5\sigma$ tension with {\it Planck}. This tension is actually still compatible with a statistical fluctuation \cite{Nunes:2021ipq}, although the first hints for a tension between the $\Lambda$CDM and the large-scale structure (LSS) data appeared already almost one decade ago \cite{Macaulay:2013swa} and have persisted over the years \cite{Gil-Marin:2016wya,Gomez-Valent:2017idt,Nesseris:2017vor,Gomez-Valent:2018nib}.

These tensions, if not explained by unaccounted systematic uncertainties in the data, might be calling for new physics to reestablish the cosmic concordance. Many models have been already invoked to try to alleviate them, with more or less success, cf. the dedicated reviews \cite{Perivolaropoulos:2021jda,DiValentino:2021izs}. Some of the most promising proposals modify the physics in the pre-recomination epoch either by considering changes in the gravitational theory \cite{SolaPeracaula:2019zsl,Ballesteros:2020sik,Braglia:2020iik,SolaPeracaula:2020vpg,Braglia:2020auw,Benevento:2022cql}, an early dark energy (EDE) component \cite{Poulin:2018cxd,Niedermann:2019olb,Ye:2020btb,Gomez-Valent:2021cbe,Wang:2022jpo}, the effect of primordial magnetic fields \cite{Jedamzik:2020krr}, varying atomic constants \cite{Liu:2019awo,Sekiguchi:2020teg} or running vacuum models \cite{SolaPeracaula:2021gxi}. These changes are usually accompanied also by some features in the post-recombination era, and the fitting performance of the various proposals can vary significantly depending on the particular data set under consideration.

In this paper we study some cosmological models with a dark energy (DE) scalar field in interaction with dark matter, in which the DE does not only play a role in the late-time universe, but can also leave some imprints in the radiation- and matter-dominated epochs (RDE and MDE, respectively). These models fall into the category of early dark energy models. While EDE is already present in standard coupled DE scenarios \cite{Wetterich:1994bg,Amendola:1999er}, here we consider the effect of scalar field potentials that have a direct impact on the universe's dynamics already in the RDE and/or MDE. Recent investigations of uncoupled EDE scenarios with a non-negligible peak of the EDE fraction before recombination have proved to be efficient in loosening the Hubble tension \cite{Poulin:2018cxd,Poulin:2021bjr,Hill:2021yec,Smith:2022hwi}, while respecting the tight upper bounds on the EDE fraction at the decoupling time \cite{Gomez-Valent:2021cbe}. This is because they are able to increase the energy budget of the universe in that era and, hence, decrease the sound horizon at the baryon-drag epoch, which forces a larger Hubble rate at low redshifts to explain the correct location of the CMB peaks and the BAO data. Some authors have pointed out, though, that these models might worsen the tension with the data on LSS, see e.g. \cite{Hill:2020osr,Gomez-Valent:2021cbe}. There is still no clear consensus on these matters (see \cite{Smith:2020rxx,Murgia:2020ryi}), and further research is needed to shed some light on the problem at hand. In particular, in this work we address the question whether the LSS estimators $\sigma_8/S_8$ are appropriate for quantifying the tension with the LSS data. We will see that, as firstly argued in \cite{Sanchez:2020vvb}, they could lead to biased results when the posterior distribution for $H_0$ is significantly different from the $\Lambda$CDM one. If so, this can have of course a direct impact on our understanding of the cosmological tensions,  as we already showed in \cite{Gomez-Valent:2021cbe}. The LSS tension could be directly intertwined with the Hubble tension. This issue has been discussed in the context of the ultra-light axion-like (ULA) EDE in \cite{Gomez-Valent:2022hkb}, where the author concluded that actually this model is able to alleviate the Hubble tension while keeping the mass fluctuations at scales of $12$ Mpc very close to the preferred value in the $\Lambda$CDM and, hence, without any clear enhancement of the clustering at these scales. Significant differences at other scales might exist, though. ULA typically enhances the power spectrum at small scales, and suppresses it at large scales. In this paper we will elaborate more on this idea. 

We concluded in our previous work \cite{Gomez-Valent:2021cbe} that a coupling between DE and DM could help us, in principle, to alleviate both tensions at a time, since it could allow the dark matter energy density to be larger than in the standard model at the last-scattering surface (which is needed to counteract the enhancement of the early integrated Sachs-Wolfe effect \cite{Vagnozzi:2021gjh}) while retrieving standard values of $\omega_{\rm cdm}$. This possibility has been already explored in the literature \cite{Karwal:2021vpk,McDonough:2021pdg}, where the authors obtained constraints using data on CMB and other data sets, incorporating in all cases the prior on $H_0$ from SH0ES \cite{Riess:2021jrx}. Here we perform our fitting analyses sticking to the combination CMB+SNIa+BAO, which contains the main ingredients to build the inverse distance ladder \cite{Aubourg:2014yra,Cuesta:2014asa,Feeney:2018mkj,Camarena:2019rmj}. We do not include any prior on the Hubble constant. We deem this important in order to assess the real ability of these models to alleviate the tensions by checking what is the actual room for shifts in the parameters of interest according to this robust cosmological data set. We provide constraints not only on $\sigma_8/S_8$, but also on $\sigma_{12}$ and the related quantity $S_{12}$ (see \cite{Sanchez:2020vvb}), and discuss the Hubble tension also in terms of the absolute magnitude of SNIa. Constraints on these parameters have not been previously reported for the coupled and uncoupled EDE models explored in this paper.

We study ULA with and without coupling in the dark sector, but not only. In \cite{Gomez-Valent:2021cbe} we showed that EDE models with scaling solutions during the matter- and radiation-dominated eras, as those found when the scalar field dynamics is governed by an exponential potential in the MDE and RDE, are not efficient concerning the tensions. Here we consider the case of DE with an exponential potential during the MDE and RDE with and without coupling with DM, and with a cosmological constant triggering acceleration at late times. We present constraints on it, and compare our results with the $\Lambda$CDM, the uncoupled model, and the coupled model with a trivial flat potential, together with the coupled and uncoupled ULA. 

On top of that, we study the effect of massive neutrinos in our analyses of uncoupled ULA and coupled quintessence with a cosmological constant. It is very well known that massive neutrinos suppress the matter power spectrum at low scales, see e.g. \cite{Lesgourgues:2006nd}, so they could compensate some of the enhancement effects found in ULA when a minimal setup with only one massive neutrino of 0.06 eV is considered, and also those introduced by the fifth force in coupled dark energy models. Understanding up to what extent this is possible is one of the main goals of this work.

The paper is organized as follows. In Sec. \ref{sec:eqs} we present the main cosmological equations at the background and linear perturbations level for the coupled dark energy models studied in this work. In Sec. \ref{sec:phenome} we discuss the most relevant phenomenological aspects of these models in order to understand their main signatures in the cosmological observables. In Sec. \ref{sec:data} we describe the data sets and the methodology employed to constrain the models, and in Sec. \ref{sec:results} we present and discuss our results. Our conclusions are provided in Sec. \ref{sec:conclusions}.


\section{Cosmological equations in coupled dark energy models}\label{sec:eqs}

We consider a perturbed flat Friedmann-Lema\^itre-Robertson-Walker universe with the spacetime line element written in the synchronous gauge,  

\begin{equation}
ds^2=a^2(\tau)[-d\tau^2+\{\delta_{ij}+h_{ij}(\tau,\vec{x})\}dx^idx^j]\,,
\end{equation}
with $\tau$ the conformal time and $\vec{x}$ the spatial comoving vector. The two scalar degrees of freedom contained in the matrix $h_{ij}$, 

\begin{equation}
h_{ij} = \int d^3k\, e^{-i\overrightarrow{k}\cdot\overrightarrow{x}}\left[h\,\hat{k}_i\hat{k}_j+6\eta\left(\hat{k}_i\hat{k}_j-\frac{\delta_{ij}}{3}\right)\right]\,,
\end{equation}
are its trace $h(\tau,\vec{k})$ and $\eta(\tau,\vec{k})$, with $\vec{k}$ the comoving wavenumber \cite{Ma:1995ey}. We do not discuss here the vector and tensor perturbations.

Regarding the matter content in the universe, we consider the fields described in the standard model of particle physics and some extension accounting for the neutrino masses, cold dark matter and dark energy. DM is treated as a pressureless perfect fluid, whereas DE is described in terms of a scalar field $\phi$ with an associated potential $V(\phi)$. We consider these two components to be in interaction. The cross-talk between DM and DE modifies their conservation equations, which take the following form, 

\begin{equation}\label{eq:ConsLaw}
 \nabla^\mu T^{\phi}_{\mu\nu}= +Q_\nu\qquad ;\qquad \nabla^\mu T^{\rm dm}_{\mu\nu}= -Q_\nu\,.
\end{equation}
The source vector $Q_\nu$ is determined by the concrete nature of the interaction. In this work we consider the conformal coupling, namely

\begin{equation}
Q_\nu = \kappa \beta T^{\rm dm}\nabla_\nu\phi\,,
\end{equation}
with $\kappa=\sqrt{8\pi G}=m_{P}^{-1}$ the inverse of the reduced Planck mass and $T^{\rm dm}$ the trace of the DM energy-momentum tensor. We use, for simplicity, a constant dimensionless coupling $\beta$. The 0-component of \eqref{eq:ConsLaw} leads to a modified equation describing the anomalous dilution law for DM, 

\begin{equation}\label{eq:DMdensity}                                                      
\bar{\rho}^\prime_{\rm dm}+3\mathcal{H}\bar{\rho}_{\rm dm} = -\beta\kappa\bar{\rho}_{\rm dm}\phi^\prime\,,
\end{equation}
and the modified Klein-Gordon (KG) equation for the scalar field, 

\begin{equation}\label{eq:KG}
\phi^{\prime\prime}+2\mathcal{H}\phi^\prime+a^2\frac{d V}{d\phi} = \beta\kappa a^2\bar{\rho}_{\rm dm}\,.
\end{equation}
The primes denote derivatives with respect to the conformal time and $\mathcal{H}=a^\prime/a$. This two equations are valid at the background level, with $\phi$ the mean (background) value of the scalar field and $\bar{\rho}_{\rm dm}$ the background DM energy density. Eq. \eqref{eq:DMdensity} can be trivially solved in terms of $\phi$. If we assume that the number of DM particles is conserved throughout the cosmic history the corresponding number density reads $n_{\rm dm}(a)=n_{\rm dm}^{(0)}a^{-3}$ and we find that their mass evolves according to

\begin{equation}\label{eq:DMmass}
m_{\rm dm}(\phi) =m_{\rm dm}^{(0)}e^{\beta\kappa(\phi^{(0)}-\phi)}\,,
\end{equation} 
with the superscripts $(0)$ denoting current quantities. The dynamics of $\phi$ is governed by eq. \eqref{eq:KG}, and depends, of course, on the intensity of the interaction and the particular shape of the scalar field potential. These features characterize the model and determine the evolution of the DM mass with the expansion as well. The Friedmann and pressure equations take the same form as in the $\Lambda$CDM, but substituting the constant dark energy density $\rho_\Lambda=\Lambda/\kappa^2$ and pressure $p_\Lambda=-\rho_\Lambda$ by 

\begin{equation}
\bar{\rho}_\phi=\frac{(\phi^\prime)^2}{2a^2}+V(\phi)\qquad {\rm and} \qquad \bar{p}_\phi=\frac{(\phi^\prime)^2}{2a^2}-V(\phi)\,. 
\end{equation}
At the linear perturbations level, and following the notation of \cite{Ma:1995ey}, we obtain in momentum space the following set of coupled differential equations \cite{Amendola:1999dr}: 

\begin{equation}\label{eq:pertFried}
\mathcal{H}h^\prime-2\eta k^2 = \kappa^2a^2\left[\delta\rho+\frac{\phi^\prime}{a^2}\delta \phi^\prime+\frac{dV}{d\phi}\delta\phi\right]\,,
\end{equation}

\begin{equation}
2\eta^\prime k^2 = \kappa^2[a^2(\bar{\rho}+\bar{p})\theta+k^2\phi^\prime\delta\phi]\,,
\end{equation}

\begin{equation}
-h^{\prime\prime}-2\mathcal{H}h^\prime+2\eta k^2 = 3\kappa^2a^2 \left[\delta p+\frac{\phi^\prime}{a^2}\delta \phi^\prime-\frac{dV}{d\phi}\delta\phi\right]\,,
\end{equation}

\begin{equation}\label{eq:pertKG}
\kappa\bar{\rho}_{\rm dm} a^2 \beta\delta_{\rm dm}=\delta\phi^{\prime\prime}+2\mathcal{H}\delta\phi^\prime+\left(k^2+a^2\frac{d^2V}{d\phi^2}\right)\delta\phi+\frac{h^\prime}{2}\phi^\prime\,,
\end{equation}

\begin{equation}
\delta_{\rm dm }^\prime = -\left(\theta_{\rm dm}+\frac{h^\prime}{2}\right)-\kappa\beta\delta\phi^\prime \,,
\end{equation}

\begin{equation}
0=\theta_{\rm dm }^\prime+(\mathcal{H}-\beta\kappa\phi^\prime)\theta_{\rm dm}+k^2\beta\kappa\delta\phi\,.
\end{equation}
The first three correspond to the 00, 0i and ii perturbed Einstein equations, respectively. The fourth one is the perturbed KG equation, and the last two are obtained from the perturbed 0 and i components of the conservation equation for dark matter. The conservation equations for the other species remain the same as in the standard model at all orders in perturbation theory, since there is no direct coupling between them and $\phi$. The quantities $\delta\rho\equiv\sum_j\delta\rho_j$ and $\delta p\equiv\sum_j\delta p_j$ are the sum of the perturbed densities and pressures of the various cosmological components, without including the scalar field contribution; $\delta_{\rm dm}$ and $\theta_{\rm dm}$ are the DM density contrast and velocity gradient, respectively; $\delta\phi$ is the perturbation of the scalar field; and $(\bar{p}+\bar{\rho})\theta\equiv \sum_j (\bar{p}_j+\bar{\rho}_j)\theta_j$. We use adiabatic initial conditions.

In order to see how the coupling impacts the large-scale structure formation processes it is illustrative to obtain the equations for the baryon and dark matter density contrasts at deep subhorizon scales in the matter- and DE-dominated universe, when radiation can be safely neglected. They read 

\begin{widetext}
\begin{equation}\label{eq:densityContrastB}
\delta_{\rm b}^{\prime\prime}+\mathcal{H}\delta^\prime_{\rm b}-\frac{\kappa^2a^2}{2}[\bar{\rho_b}\delta_b+\bar{\rho}_{\rm dm}\delta_{\rm dm}]=0\,,
\end{equation}
\begin{equation}\label{eq:densityContrastDM}
\delta_{\rm dm}^{\prime\prime}+\left[\mathcal{H}-\frac{\beta\kappa\phi^\prime k^2}{k^2+a^2m_\phi^2}\right]\delta^\prime_{\rm dm}+(\delta_b^\prime-\delta^\prime_{\rm dm})\frac{\beta\kappa\phi^\prime a^2m_\phi^2}{k^2+a^2m_\phi^2}-\frac{\kappa^2a^2}{2}\left[\bar{\rho}_b\delta_b+\bar{\rho}_{\rm dm}\delta_{\rm dm}\left(1+\frac{2\beta^2k^2}{k^2+a^2m^2_{\phi}}\right)\right]=0\,.
\end{equation}
\end{widetext}
These equations encode the non-trivial growth of the matter perturbations, which depends a lot on the details of the model under study. Notice that we have also included the effect of the DE mass $m^2_{\phi}(a)\equiv d^2V/d\phi^2$. A larger mass confines the fifth force effects to smaller scales, i.e. to larger $k$'s. Nevertheless, the DE potentials that are able to trigger the late-time acceleration of the universe must be extremely flat at late times to explain the current data, see e.g. \cite{Planck:2018vyg,SolaPeracaula:2018wwm}. Hence, $m_\phi$ can be neglected at low redshifts. However, if the shape of $V(\phi)$ allows for a non-negligible effective DE mass in previous stages of the cosmic expansion, competing with the physical wave modes $k/a$ inside the horizon, this could leave an imprint on the large-scale structure of the universe, so it is useful to keep the mass terms in these equations and discuss their effects on a case-by-case basis.  

It is easy to check that in the limit $\beta\to 0$ of Eqs. \eqref{eq:densityContrastB}-\eqref{eq:densityContrastDM} we recover the results of the $\Lambda$CDM and uncoupled dark energy models, and in the limit $m_\phi\to 0$ we retrieve the well-known result of coupled dark energy with a nearly flat potential, see e.g. \cite{Barros:2018efl,Gomez-Valent:2020mqn},

\begin{equation}\label{eq:densityContrastDM2}
\delta_{\rm dm}^{\prime\prime}+\left[\mathcal{H}-\beta\kappa\phi^\prime\right]\delta^\prime_{\rm dm}-\frac{\kappa^2a^2}{2}\left[\bar{\rho}_b\delta_b+\bar{\rho}_{\rm dm}\delta_{\rm dm}(1+2\beta^2)\right]=0\,.
\end{equation}
By combining \eqref{eq:pertFried} and \eqref{eq:pertKG} one can see that in these coupled DE scenarios the dark energy component does not cluster at deep subhorizon scales, since when $k^2\gg\mathcal{H}^2$ we have 

\begin{equation}
\delta\rho_\phi=\frac{\phi^\prime}{a^2}\delta \phi^\prime+\frac{dV}{d\phi}\delta\phi\sim \left(\frac{H}{k}\right)^2\times\left[\beta^2,\beta\alpha\right]\times\mathcal{O}(\delta\rho_m)\,,
\end{equation}
with $\alpha$ the slope of the potential. This means that galaxies are tracers of the underlying distribution of dark matter, as in the $\Lambda$CDM.

In this work we study models with a potential of the following form, 

\begin{equation}
V(\phi)=V_0+V_{\rm ede}(\phi)\,.
\end{equation}
By construction, $V_{\rm ede}$ is not important at $z\lesssim 1$, but can have a sizable impact at $z\gg 1$, i.e. during the matter and/or radiation-dominated epochs. $V_0$, instead, is a constant term that is responsible of the current accelerated phase of the universe. Therefore, $V_0=\mathcal{O}(m_{P}^2H_0^2)$ and dominates the expansion at late times. It is basically the energy density associated to the cosmological constant. For simplicity we do not consider more complicated shapes of the potential at low redshifts, just not to introduce possible degeneracies with the features that characterize $V_{\rm ede}$. This would hinder the interpretation of our results.

We focus in this paper on three alternative forms of $V_{\rm ede}$ with and without considering the effect of the coupling in the dark sector, i.e. studying both the cases with $\beta=0$ and $\beta\ne 0$. We dedicate the next section to list them and explain their main phenomenological aspects.


\section{Phenomenology of several coupled and uncoupled EDE models in a nutshell}\label{sec:phenome}

\subsection{Constant potential}\label{sec:constPot}

\noindent {\it Standard model:} If $V(\phi)=V_0$ and $\beta=0$ the model reduces to the $\Lambda$CDM, since in this case the DE has no dynamics and there is no interaction between DM and DE. This is the only model studied in this paper in which the dark energy density is a pure rigid constant at all times. 
\newline
\newline
{\it Coupled dark energy with a flat potential:} If $V(\phi)=V_0$ but $\beta\ne 0$ the interaction in the dark sector is active, and this gives rise to a very rich phenomenology. The first coupled DE models of this sort where presented and studied in \cite{Wetterich:1994bg,Amendola:1999er}. See also \cite{AmendolaTsujikawaBook}. In this paper we refer to the model of coupled DE with flat potential as CDE\_const. For values of the coupling of the order $\beta\lesssim \mathcal{O}(10^{-1})$, as the ones preferred by the data \cite{Amendola:2000ub,Pettorino:2012ts,Pettorino:2013oxa,Xia:2013nua,Planck:2015bue,vandeBruck:2016hpz,vandeBruck:2017idm,Barros:2018efl,Agrawal:2019dlm,Gomez-Valent:2020mqn,Gomez-Valent:2022hkb,Goh:2022aaa}, the scalar field has a negligible energy density before the matter-radiation equality time, $t_{\rm eq}$. In particular, its kinetic energy is derisory, and we can take as initial condition $\phi^\prime_{\rm ini}\equiv \phi^{\prime}(z_{\rm ini})=0$, with $z_{\rm ini}=10^{14}$. In this model only the derivatives of the scalar field enter the equations, so the cosmological evolution of the relevant quantities is not affected by the initial value of $\phi$. We opt to set $\phi_{\rm ini}=0$. The interaction starts to play a role around $t_{\rm eq}$, when the value of $\bar{\rho}_{\rm dm}$ becomes comparable to the one of the radiation energy density and the source term in the {\it rhs} of the KG equation \eqref{eq:KG} starts to accelerate the scalar field. In this model the product $\beta\phi^\prime>0$ regardless of the sign of $\beta$, i.e. the {\it rhs} of eq. \eqref{eq:DMdensity} is negative, so there is a decay of the DM mass that happens mostly during the matter-dominated era. The model has an almost exact scaling solution in that epoch, with an EDE fraction $\Omega^{\rm MD}_{\rm ede}\approx 2\beta^2/3$ \cite{Amendola:1999er}\footnote{It would be exact if also baryons were coupled to the DE. This is a possibility that we do not contemplate in this paper, just to automatically pass the stringent local constraints on fifth forces.}. Thus, there is a non-zero amount of kinetic EDE during the post-recombination epoch, even when the potential is a simple constant, as in the case we are studying here. An increase of the absolute value of the coupling enhances the fifth force and, for fixed initial conditions, it makes the universe to be more decelerated and DM to cluster more efficiently in the matter-dominated epoch. If we consider that baryons are also coupled to DE as a first approximation, the deceleration parameter and density contrast read, respectively, $q=-\ddot{a}a/\dot{a}^2=\frac{1}{2}+\beta^2$ and $\delta_{\rm m}\sim a^{1+2\beta^2}$.

\begin{table}[t!]
\centering
\begin{tabular}{|c || c|   }
\hline
{\small Shorthand}  & {Model} 
\\\hline
{\small CDE\_const} & Coupled DE with constant potential 
\\\hline
{\small EXP} &  Uncoupled DE with exponential potential
\\\hline
{\small CDE\_EXP} &  Coupled DE with exponential potential
\\\hline
{\small ULA} &  Uncoupled DE with ULA potential
\\\hline
{\small CDE\_ULA} & Coupled DE with ULA potential 
\\\hline
\end{tabular}
\caption{Shorthands for the different EDE models studied in this paper.}
\label{tab:shorthands}
\end{table}

A larger mass of the DM particles at the last scattering surface could help, in principle, to alleviate the Hubble tension, since this would decrease the sound horizon at the baryon-drag epoch, $r_d$, which would require in turn a larger Hubble function at late times to keep fixed the location of the first peak of the CMB temperature power spectrum measured by {\it Planck}. In practice, though, current data put very tight constraints on $\beta$ and limit strongly the capability of the model of loosening the $H_0$ tension \cite{Gomez-Valent:2020mqn}\footnote{Weaker constraints on $\beta$ are obtained when the low multiples from {\it Planck} are combined with ACT and SPT-3G CMB data, which allows for larger values of  $H_0$. However, the alleviation of the Hubble tension is in this case induced by a reduction of the statistical power of the overall data set \cite{Goh:2022aaa}.}. Nevertheless, past studies found a persistent peak in the posterior distribution of the coupling under various data sets and with different statistical significance, which is certainly an interesting feature of the model, see e.g. \cite{Pettorino:2013oxa,Planck:2015bue,Gomez-Valent:2020mqn}. This peak has recently been shown not to be induced by volume effects introduced in the marginalization process \cite{Gomez-Valent:2022hkb}.

We deem useful to present in this work the fitting results for the $\Lambda$CDM and coupled DE with a constant potential and compare them with those obtained with the coupled and uncoupled EDE models that we describe in the next two subsections, \ref{sec:expPot} and \ref{sec:ULAPot}.


\subsection{Exponential potential}\label{sec:expPot}

We also study the case of EDE with the following exponential scalar field potential

\begin{equation}\label{eq:expPot}
V(\phi)=V_0+\frac{\Omega^{\rm RD}_{\rm ede}\rho_r(a_{\rm ini})}{3(1-\Omega^{\rm RD}_{\rm ede})}\exp\left[\frac{-2\kappa}{\sqrt{\Omega^{\rm RD}_{\rm ede}}}(\phi-\phi_{\rm ini})\right]\,.
\end{equation}
During the RDE the model has a scaling solution with an EDE fraction equal to $\Omega^{\rm RD}_{\rm ede}$ \cite{Wetterich:1987fm,Copeland:1997et}. This parameter is left free in our Monte Carlo analyses, with the flat prior $\Omega^{\rm RD}_{\rm ede}\in [0,1]$. It is the only additional parameter with respect to the $\Lambda$CDM. The constant factor that is multiplying the exponential term in formula \eqref{eq:expPot} has been chosen to make the model to be already in the scaling regime at $z_{\rm ini}$, and $\phi_{\rm ini}$ can be safely set to $0$, since the dynamics of the scalar field depends only on the difference $\phi-\phi_{\rm ini}$ \footnote{ Notice that the scaling regime in the MDE and RDE would be never reached if the constant factor $\lambda$ in the exponential $e^{-\lambda\kappa\phi}$ was $\lambda<\sqrt{2}$ \cite{Copeland:1997et,Amendola:1999er}. In the latter case the exponential potential would only become important at low redshifts. Values of  $\lambda<\sqrt{2}$ are required to produce the late-time accelerated phase of the universe if $V_0=0$, of course. Constraints on the model with $V_0=0$ and an exponential potential with $\lambda<\sqrt{2}$ (and other similar potentials) have been obtained in many previous works in the literature, see again \cite{Amendola:2000ub,Pettorino:2012ts,Pettorino:2013oxa,Xia:2013nua,Planck:2015bue,vandeBruck:2016hpz,vandeBruck:2017idm,Barros:2018efl,Agrawal:2019dlm,Gomez-Valent:2020mqn,Gomez-Valent:2022hkb}. We remark that the model CDE\_EXP is very different from the latter, since we automatically have $\lambda>\sqrt{2}$ due to the physical range of values allowed for $\Omega_{\rm ede}^{\rm RD}$. The constant term in the potential \eqref{eq:expPot}, $V_0$, is crucial to ensure the phenomenological viability of the model.}. 

\begin{figure}[t!]
\begin{center}
\includegraphics[width=2.5in, height=2in]{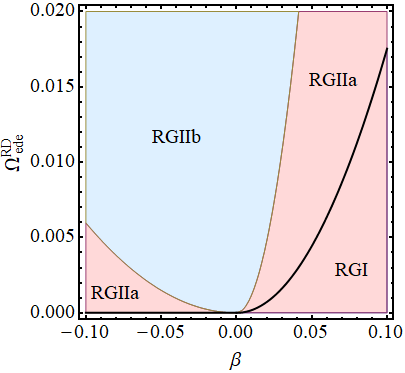}
\caption{Diagram in the plane $(\beta,\Omega_{\rm ede}^{\rm RD})$ of  CDE\_EXP. The black thick line is the border between regions RGI and RGII. Below it and in the line $\Omega_{\rm ede}^{\rm RD}=0$ we have region RGI \eqref{eq:condlower}, whereas above it we have RGII \eqref{eq:condlarger}. The blue part corresponds to the region of parameter space that leads to a lower growth of matter fluctuations than the $\Lambda$CDM during the scaling regime in the MDE, $\delta\sim a^n$ with $n<1$, whereas the pink one indicates the region at which $n>1$. The border between the latter is set by the curves $\Omega_{\rm ede}^{\rm RD}=4\beta^2/(-1+\sqrt{5/2})^2+\mathcal{O}(4)$ if $\beta\geq 0$ and $\Omega_{\rm ede}^{\rm RD}=4\beta^2/(1+\sqrt{5/2})^2+\mathcal{O}(4)$ if $\beta<0$. See Sec. \ref{sec:expPot} for details.}\label{fig:fig1}
\end{center}
\end{figure}

In the MDE the model has two fixed points. The model is attracted by one or the other depending on the values of $\Omega^{\rm RD}_{\rm ede}$ and the coupling $\beta$ \cite{Amendola:1999er}. These fixed points are not stable due to the non-zero constant term $V_0$ in the potential, which forces the model to abandon the scaling regime when matter is sufficiently diluted. The universe enters then the phase of late-time acceleration.  Let us analyze first the phenomenology in the case when there is no coupling in the dark sector.
\newline
\newline
{\it Exponential potential with no coupling:} We have $\Omega^{\rm RD}_{\rm ede}\ne 0$, but $\beta=0$. We call this model ``EXP''. It is a particular case of the double exponential potential firstly studied in \cite{Barreiro:1999zs}. The EDE fraction in the MDE reads in this case $\Omega_{\rm ede}^{\rm MD}=3\Omega_{\rm ede}^{\rm RD}/4$, i.e. it decreases by $25\%$ from the RDE to the MDE \cite{Wetterich:1987fm,Copeland:1997et}. EDE delays the matter-radiation equality time and slows the growth of matter perturbations down with respect to the $\Lambda$CDM. During the MDE the matter density contrast at subhorizon scales grows as $\delta_m\sim a^{1-\frac{9}{20}\Omega^{\rm RD}_{\rm ede}}$. The presence of DE near last scattering affects the relation between the position of CMB peaks \cite{Doran:2000jt,Doran:2001yw}. The maximum value of $\Omega^{\rm MD}_{\rm ede}$ is actually extremely constrained by the CMB data, basically due to the stringent bounds imposed on the EDE fraction at the last scattering surface. We found in \cite{Gomez-Valent:2021cbe} an upper bound $\Omega^{\rm MD}_{\rm ede}\lesssim 0.3\%$ at $1\sigma$ c.l. using the {\it Planck} CMB data and the supernovae of Type Ia (SNIa) from the Pantheon compilation. This tight constraints render the model extremely close to the $\Lambda$CDM, in practice, so its impact on the cosmological tensions is small. Here we also provide constraints on this model to ease the comparison with the results obtained when we take into account the effect of the coupling.  
\newline
\newline
{\it Coupled dark energy with an exponential potential:} We call this model ``CDE\_EXP'' along this paper. Here we consider the most general scenario, with both $\Omega^{\rm RD}_{\rm ede}\ne 0$ and $\beta\ne 0$. The concrete scaling solution during the MDE is usually reached at $z\sim \mathcal{O}(100)$ and depends, as mentioned before, on the values of these two parameters \cite{Amendola:1999er}. 

\begin{figure}[t!]
\begin{center}
\includegraphics[width=3.3in, height=4.24in]{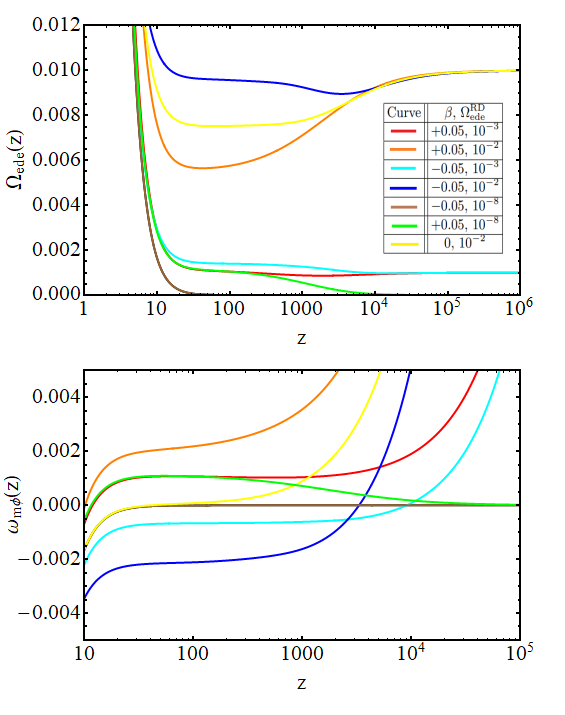}
\caption{{\it Upper plot}: EDE fraction as a function of the redshift for the CDE\_EXP models listed in the legend; {\it Lower plot:} EoS parameter of the matter+DE fluid. During the MDE $\Omega_{\rm ede}(z)$ and $w_{\rm m\phi}(z)$ tend to the scaling solutions discussed in the main text.}\label{fig:fig2}
\end{center}
\end{figure}

In region 

\begin{equation}\label{eq:condlower}
{\rm RGI:}\quad\beta>0\qquad {\rm and}\quad \Omega_{\rm ede}^{\rm RD}<\frac{16\beta^2}{(3+2\beta^2)^2}
\end{equation}
the EDE fraction and deceleration parameter in the scaling regime read $\Omega_{\rm ede}^{\rm MD}\approx 2\beta^2/3$ and $q\approx 1/2+\beta^2$, respectively, and the equation of state (EoS) of the composite matter+DE fluid $w_{\rm m\phi}\equiv {\bar p}_\phi/(\bar{\rho}_{\rm m}+\bar{\rho}_\phi)\approx 2\beta^2/3$. The DM density contrast $\delta_m\sim a^{1+2\beta^2}$ at subhorizon scales, as it happens also in the model CDE\_const described in Sec. \ref{sec:constPot}. It is important to remark, though, that in this model the growth of perturbations can be very different to the one found in CDE\_const in previous stages of the cosmic history due to the effect of the non-trivial EDE potential \eqref{eq:expPot}. We will discuss this explicitly later on. The exponential potential in this model does not play any role during the scaling regime, since DE only has kinetic energy.

Instead, in region

\begin{equation}\label{eq:condlarger}
{\rm RGII:}\quad \beta\leq 0\qquad {\rm or} \quad \Omega_{\rm ede}^{\rm RD}>\frac{16\beta^2}{(3+2\beta^2)^2}\,,
\end{equation} 
we have the following scaling formulas

\begin{equation}
\Omega_{\rm ede}^{\rm MD}\approx \frac{3}{4}\Omega_{\rm ede}^{\rm RD}-\frac{\beta}{2}\sqrt{\Omega_{\rm ede}^{\rm RD}}+\mathcal{O}(4)\,,
\end{equation}
\begin{equation}
q\approx\frac{1}{2}+\frac{3}{4}\beta\sqrt{\Omega_{\rm ede}^{\rm RD}}+\mathcal{O}(4)\,,
\end{equation}
\begin{equation}
w_{\rm m\phi}\approx\frac{\beta}{2}\sqrt{\Omega_{\rm ede}^{\rm RD}}+\mathcal{O}(4)\,,   
\end{equation}
where $\mathcal{O}(4)\equiv \mathcal{O}(\beta^2\Omega_{\rm ede}^{\rm RD},\beta(\Omega_{\rm ede}^{\rm RD})^{3/2})$, i.e. we consider $\beta\sim\mathcal{O}(1)$ and $\sqrt{\Omega_{\rm ede}^{\rm RD}}\sim\mathcal{O}(1)$. Negative values of $\beta$ produce larger EDE fractions and a less decelerated universe during the scaling period, contrary to what we find when $\Omega_{\rm ede}^{\rm RD}=0$, i.e. in CDE\_const. In this epoch the growing mode of the matter density contrast at subhorizon scales evolves as $\delta_m\sim a^n$, with

\begin{equation}\label{eq:power}
n=1+\frac{6}{5}\beta^2+\frac{6}{5}\beta\sqrt{\Omega_{\rm ede}^{\rm RD}}-\frac{9}{20}\Omega_{\rm ede}^{\rm RD}+\mathcal{O}(4)\,.
\end{equation}
Notice that in this case there exists a region in the plane $(\beta,\Omega_{\rm ede}^{\rm RD})$ that leads to values of $n<1$, i.e. that lets matter perturbations to grow slower than in the standard model. We call this region RGIIb to distinguish it from the region with an enhanced growth, RGIIa. Of course, all these expressions reduce to the ones found under the condition \eqref{eq:condlower} when $\beta>0$ and $\Omega_{\rm ede}^{\rm RD}\to 16\beta^2/9+\mathcal{O}(4)$. Moreover, as expected, we also retrieve the expressions for the model EXP in the limit $\beta\to 0$. The limit $\Omega_{\rm ede}^{\rm RD}\to 0$ is more subtle and counterintuitive. One would expect to recover the scaling solutions of the model CDE\_const, but this only happens if we perform the limit in the region RGI. In RGIIa this limit leads to different results. In the presence of $\Omega_{\rm ede}^{\rm RD}\ne0$, regardless how small it is, there is a clear asymmetry between positive and negative values of the coupling, which is not present when $\Omega_{\rm ede}^{\rm RD}$ is exactly equal to zero. We find, for instance, that during the scaling period, for extremely small values of the EDE fraction in the RDE, $\Omega_{\rm ede}^{\rm MD}$ is negligible if $\beta<0$, whereas it is proportional to $\beta^2$ if $\beta>0$. 

\begin{figure}[t!]
\begin{center}                
\includegraphics[width=2.7in, height=2.1in]{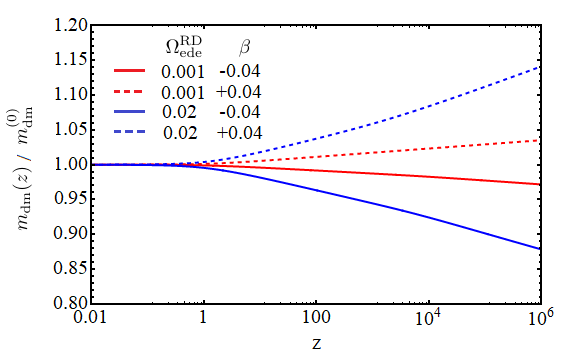}
\caption{Evolution of the mass ratio $g(z)$ \eqref{eq:gmass} in the CDE\_EXP model for four different sets of parameters $(\Omega_{\rm ede}^{\rm RD},\beta)$. The second one (red dotted line) is located in the region RGI; the others are in RGII. Notice that for $\Omega_{\rm ede}^{\rm RD}\ne 0$ if $\beta<0$ the DM mass grows with the expansion, whereas it decreases if $\beta>0$. See the main text for details.}\label{fig:fig3}
\end{center}
\end{figure}

\begin{figure*}[t!]
\begin{center} 
\includegraphics[width=7.1in, height=2.3in]{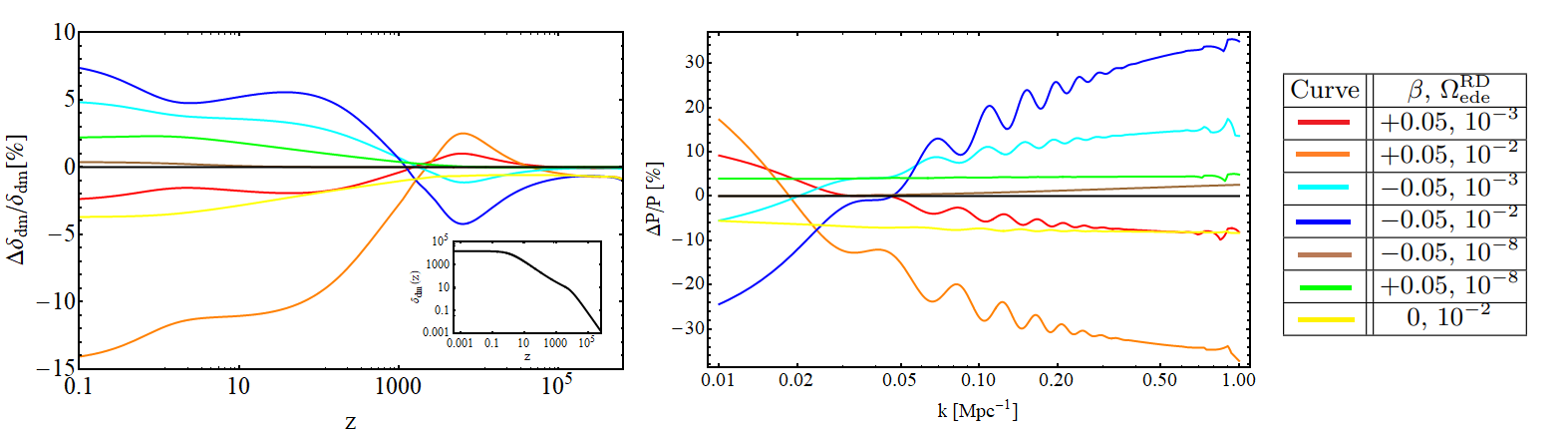}
\caption{{\it Left plot:} Relative differences between the density contrast of DM computed in the CDE\_EXP models specified in the table and the $\Lambda$CDM (our reference model), as a function of the redshift and for $k=0.1$ Mpc$^{-1}$. For the standard model we use the values of the parameters obtained from the Planck 2018 TTTEEEE analysis \cite{Planck:2018vyg}. In the inner plot we show the evolution of $\delta_{\rm dm}(z)$ in the $\Lambda$CDM, to understand when these relative differences lead to important absolute differences; {\it Right plot:} The same, but for the linear matter power spectrum at $z=0$, in the relevant range of $k$'s.}\label{fig:fig4}
\end{center}
\end{figure*}

In Fig. \ref{fig:fig1} we show a diagram of the $(\beta,\Omega_{\rm ede}^{\rm RD})$-plane, with the two regions defined by the conditions \eqref{eq:condlower} and \eqref{eq:condlarger}, and also indicating with different colors the regions that enhance or suppress the growth of matter perturbations during the scaling time in the MDE with respect to the $\Lambda$CDM when the initial conditions and the other parameters are kept fixed in the analysis. It is clear that the presence of EDE in the RDE introduces an asymmetry between the solutions found for negative and positive values of the coupling constant $\beta$, which was not present in CDE\_const. This asymmetry, though, does not generate big changes in the power $n$ \eqref{eq:power}, which satisfies $|n-1|<2\%$ for all the points in the region of parameter space shown in Fig. \ref{fig:fig1}. We will see later that, given some fixed initial conditions, the largest deviations from the $\Lambda$CDM regarding the growth of matter perturbations happen during the MDE, but before the scaling period, i.e. in the redshift range $50 \lesssim z\lesssim z_{\rm eq}$, with $z_{\rm eq}=z(t_{\rm eq})$.

We show in Fig. \ref{fig:fig2} the evolution of the EDE fraction and the EoS of the matter+DE fluid for several values of the CDE\_EXP parameters.

It is also important to understand how the DM mass evolves in this model, since it can make the radiation-equality time to happen earlier or later in the cosmic history, and has an obvious impact on the LSS as well. In contrast to model CDE\_const, in CDE\_EXP the mass of the dark matter particles does not remain constant during the RDE due to the velocity of the scalar field triggered by the potential, and can also increase throughout the history of the universe depending on the sign of $\beta$. Using Eq. \eqref{eq:DMmass} it is straightforward to show that

\begin{equation}\label{eq:gmass}
g(a)\equiv \frac{d}{da}\left(\frac{m_{\rm dm}(a)}{m_{\rm dm}^{(0)}}\right)= -\frac{m_{\rm dm}}{m^{(0)}_{\rm dm}}\left[\beta \frac{d(\kappa\phi)}{da}\right]\,.
\end{equation}
During the RDE $g_{\rm RD}\propto -\beta\sqrt{\Omega_{\rm ede}^{\rm RD}}$, and during the scaling period in the MDE

\begin{equation}
g_{\rm MD} \propto \left\{
        \begin{array}{ll}
            -\beta^2 & \quad {\rm in\,\,RGI} \\
            -\beta\sqrt{\Omega_{\rm ede}^{\rm RD}} & \quad {\rm in\,\,RGII} 
        \end{array}
    \right.
\end{equation}
Thus, the DM mass in the region RGI of parameter space decreases with the scale factor, since $\beta>0$. In region RGII, instead, it decreases if $\beta>0$ and increases if $\beta<0$. We have verified numerically that it is also the case between the RDE and the scaling period in the MDE. Hence, we find that if $\Omega_{\rm ede}^{\rm RD}\ne 0$ the DM mass grows with the expansion if $\beta<0$ and decreases if $\beta>0$. If $\Omega_{\rm ede}^{\rm RD}= 0$ (i.e. in CDE\_const) $m_{\rm dm}$ always decreases. 
In Fig. \ref{fig:fig3} we plot the evolution of the DM mass for different sets of the parameters $\Omega_{\rm ede}^{\rm RD}$ and $\beta$. It is clear that in order to keep the DM mass around the recombination time $|\Delta m_{\rm dm}/m^{(0)}|\lesssim 5\%$, with  $|\beta|\sim 0.04$, we have to demand $\Omega_{\rm ede}^{\rm RD}\lesssim \mathcal{O}(1\%)$, which is a similar upper bound to the one obtained in the uncoupled scenario. Thus, we do not expect the non-null coupling to relax a lot the existing tight constraints on $\Omega_{\rm ede}^{\rm RD}$. We will actually see in Sec. \ref{sec:results} that this is precisely the order of magnitude that we get from our fitting analyses.

Now we discuss in more detail the evolution of the DM density contrast in the CDE\_EXP model, fixing the initial conditions and exploring different values of $\beta$ and $\Omega_{\rm ede}^{\rm RD}$. In the left plot of Fig. \ref{fig:fig4} we show the relative differences between the density contrast of DM computed in the $\Lambda$CDM and the other models, as a function of the redshift. In all cases the effect of the DE mass can be neglected in good approximation, since the comoving perturbation scale under study ($k=0.1$ Mpc$^{-1}$) is larger than the product $am_\phi$ after horizon crossing, see Appendix A. The enhancement or suppression of the DM density contrast affects the evolution of the baryon density perturbations through equation \eqref{eq:densityContrastB}, and leads also to the increase or decrease of the total matter power spectrum, see the right plot in Fig. \ref{fig:fig4}.

First of all let us analyze the simplest scenario, in which the coupling is switched off and there is a non-null EDE fraction (yellow curve). Before the scale reenters the horizon at $z\sim 10^6$ the DM density contrast remains constant, $1\%$ below the $\Lambda$CDM value due to the $1\%$ additional radiation content in form of DE. Apart from this, nothing differs from the standard model, except that during the MDE the matter fraction is lower and DE is non-zero. This suppresses the perturbations growth.

\begin{figure*}[t!]
\begin{center} 
\includegraphics[width=7.2in, height=5.0in]{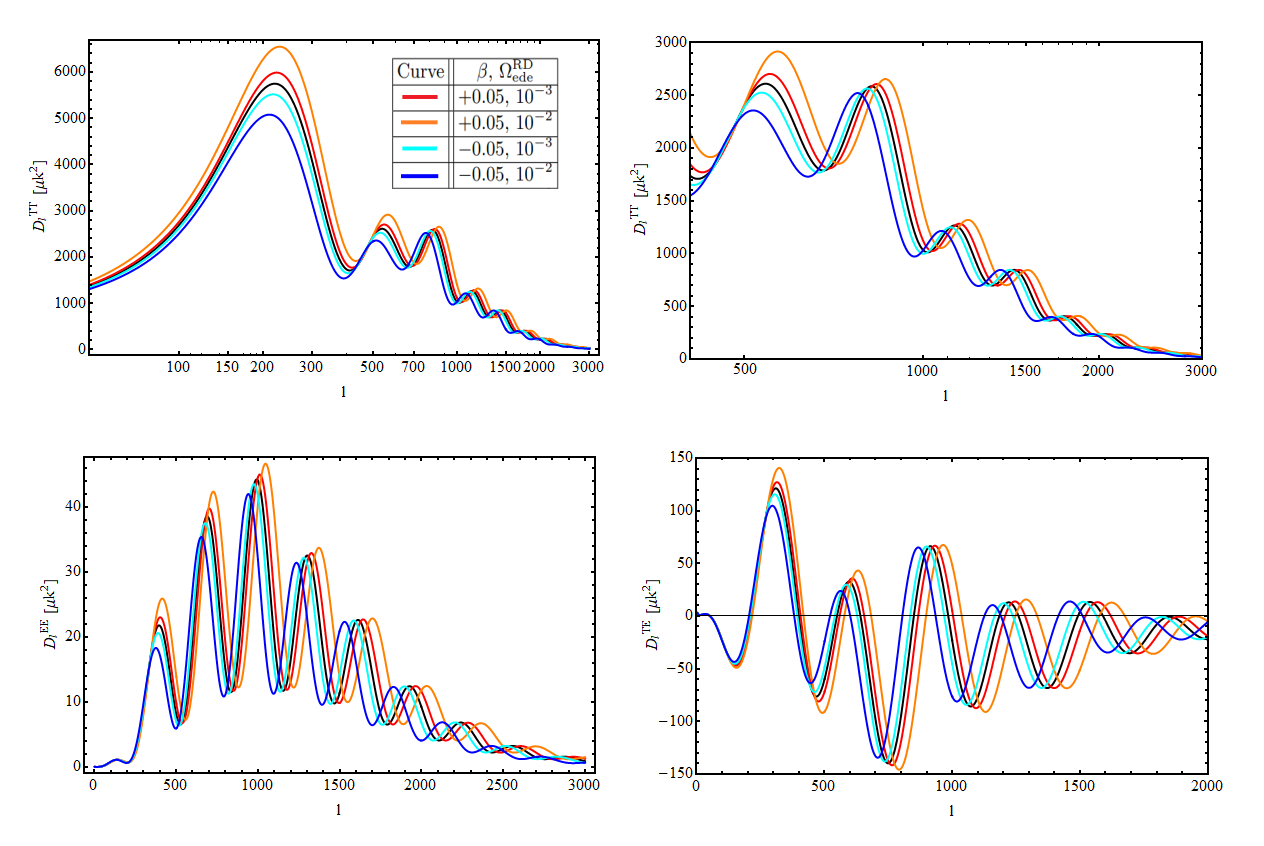}
\caption{CMB temperature (TT) and polarization (EE) auto-correlation spectra, and their cross-correlation (TE), obtained for four different CDE\_EXP scenarios. In the top-right plot we zoom in the region $l\gtrsim 500$ to better distinguish the differences between the TT spectra in that multipole range. The black curves correspond to the reference $\Lambda$CDM model, the same employed in Fig. \ref{fig:fig4}.}\label{fig:fig5}
\end{center}
\end{figure*}

What does it happen, instead, when both, $\beta$ and $\Omega_{\rm ede}^{\rm RD}$, are sizable? Once the scale reenters the horizon, before the matter-radiation equality time the matter fraction is negligible. The scalar field has been accelerated during the RDE, though, so it has some velocity and the friction term in the equation for $\delta_{\rm dm}$ \eqref{eq:densityContrastDM2} modifies the evolution of the density contrast, since it is proportional to $-\beta\phi^\prime$. If $\beta<0$ ($\beta>0$) the friction is enhanced (suppressed). These departures from the $\Lambda$CDM do not have a huge impact in terms of absolute differences, since they happen when the density perturbations are still quite small (see the inner plot in Fig. \ref{fig:fig4}). After $z_{\rm eq}$, the matter fraction in the universe starts to be important and so also the source term in Eq. \eqref{eq:densityContrastDM2}. The typical variation of $\kappa\phi$ triggered by the potential during the RDE is of order one and positive for the values employed to build the orange, red, blue and cyan curves ($\Omega_{\rm ede}^{\rm RD}=10^{-3},10^{-2}$). This statement is easy to prove. During the RDE the scalar field does not feel the coupling and evolves as 

\begin{equation}
\kappa\phi(a)=\kappa\phi_{\rm ini}+2\sqrt{\Omega_{\rm ede}^{\rm RD}}\ln\left(\frac{a}{a_{\rm ini}}\right) \,,
\end{equation}
Thus,

\begin{equation}
\kappa\phi(z_{\rm eq})-\kappa\phi_{\rm ini}\sim 50\sqrt{\Omega_{\rm ede}^{\rm RD}}\sim \mathcal{O}(1)   
\end{equation}
for the aforementioned range of values of $\Omega_{\rm ede}^{\rm RD}$. This basically means that at $z_{\rm eq}$ the mass of the DM particles \eqref{eq:DMmass} takes the value 

\begin{equation}
 m_{\rm dm}(z_{\rm eq})=m_{\rm ini}[1-\beta\times\mathcal{O}(1)]\,.
\end{equation}  
Contrary to what we find when there is no important EDE fraction in the RDE, the source term in Eq. \eqref{eq:densityContrastDM2} contains a contribution that is proportional to $\beta$, instead of $\beta^2$. For $\beta>0$ the mass is smaller than $m_{\rm ini}$ and the DM density contrast decreases with respect to the one in the $\Lambda$CDM. If the coupling is negative the opposite happens. This behavior can be clearly appreciated in Fig. \ref{fig:fig4}. When the system reaches the scaling regime, at $z\sim 50$, we can apply the formulas of $\delta_{\rm dm}$ shown before. For the values of $(\beta,\Omega_{\rm ede}^{\rm RD})$ contained in the region RGIIb (blue curve) of the parameter space there is a suppression of the DM growth during the scaling period. Nevertheless, as we have already pointed out before, the largest fraction of the relative differences are generated between the matter-radiation equality time and the beginning of the scaling in the MDE.

The brown and green curves in fig. \ref{fig:fig4} deserve also some comments. Due to the very small value of $\Omega_{\rm ede}^{\rm RD}$ in these cases, the friction term plays no role before $t_{\rm eq}$, and this is why we do not observe any bump at $z>z_{\rm eq}$ in the plot of $\Delta\delta_{\rm dm}/\delta_{\rm dm}$. The mass of the DM particles at that time is very close to the initial one, $m_{\rm ini}$, so the source term in Eq. \eqref{eq:densityContrastDM2}  receives only a correction of order $\beta^2$. Hence, regardless of the sign of $\beta$, $\delta_{\rm dm}$ starts to grow after $t_{\rm eq}$. The growth is faster if $\beta>0$, since the increase of the scalar field and the DM mass is more efficient in this case.

We illustrate in Fig. \ref{fig:fig5} the behavior of the CMB temperature and polarization spectra and their cross-correlation in the the CDE\_EXP models. For fixed initial conditions, a negative coupling makes the mass of the DM particles to increase with the expansion (see again Fig. \ref{fig:fig3}), which leads to a larger DM energy density compared to the $\Lambda$CDM. The relative difference is larger in the late-time universe and this produces an increase of the angle subtended by the sound horizon at the decoupling time. These phenomena, which are enhanced by larger values of $\Omega_{\rm ede}^{\rm RD}$, in turn, produce a decrease of the amplitude of the CMB peaks and their shift towards lower multipoles. This explains why $\beta$ is positively correlated with $H_0$.


\subsection{ULA potential}\label{sec:ULAPot}

We also want to test the performance of the coupled and uncoupled dark energy models ruled by the ultra-light axion-like potential \cite{Poulin:2018dzj,Poulin:2018cxd},

\begin{equation}\label{eq:ULApotential}
V(\phi)=V_0+m^2f^2[1-\cos(\phi/f)]^3\,.
\end{equation} 
This potential introduces three additional parameters with respect to the $\Lambda$CDM, to wit: $m$, $f$ and the initial value of the scalar field, all of them with dimensions of energy in natural units. The latter is usually written in terms of the dimensionless quantity $\theta_{\rm ini}=\phi_{\rm ini}/f$, with $\theta_{\rm ini}\in[0,\pi]$. Deep in the RDE, and regardless of the coupling, the expansion rate is much larger than the mass of $\phi$, i.e. $H\gg m$, so the scalar field has practically no dynamics and $V\simeq const.$ This allows us to take as initial condition $\phi^\prime_{\rm ini}=0$. 

\begin{figure*}[t!]
\begin{center}
\includegraphics[width=5.3in,height=3.7in]{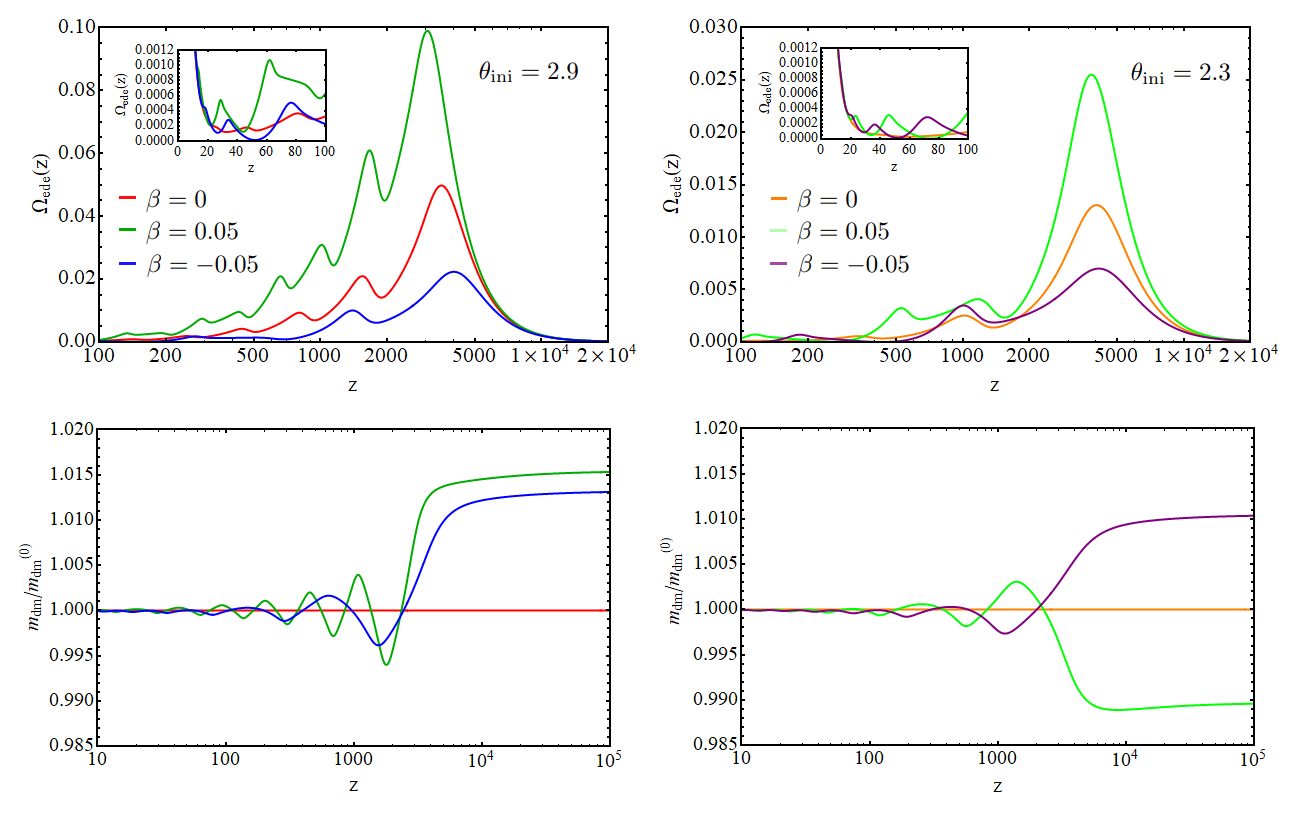}
\caption{{\it Upper plots:} EDE fractions obtained with different ULA and CDE\_ULA models. We use in all cases the same values of $\omega_b$, $V_0$, $\bar{\rho}_{\rm dm}(z_{\rm ini})$, $\log_{10}(f/{\rm eV})$ and  $\log_{10}(m/{\rm eV})$. In the left plot we set $\theta_{\rm ini}=2.9$, and in the right plot $\theta_{\rm ini}=2.3$. We show the results for $\beta=0,+0.05, -0.05$; {\it Lower plots:} The value of the mass ratio $m_{\rm dm}(z)/m_{\rm dm}^{(0)}$ for the models studied in the upper plots. Positive values of $\beta$ lead to a larger $f_{\rm ede}$, but do not necessarily make the DM mass to decrease with the expansion. For sufficiently low values of $\theta_{\rm ini}$ a $\beta>0$ produce $m_{\rm dm}^{(0)}>m_{\rm dm}(z_{\rm ini})$. Notice also that a very significant fraction of the total variation of the DM mass happens between $z_{\rm max}$ and the decoupling time.  See the main text for more comments.}\label{fig:fig6}
\end{center}
\end{figure*}

We discuss the models with and without coupling separately in the subsequent paragraphs:
\newline
\newline
{\it ULA with no coupling:} In this case $\beta=0$. We refer to this model simply as ``ULA''. The scalar field gets accelerated when $H\sim m$. At that moment it starts to roll down the potential and eventually oscillates around its minimum. During the oscillatory phase its energy density decays faster than radiation, and after the decay the model reduces essentially to the standard model, with $V(\phi)\approx V_0$. If the constant value of the potential during the RDE is large enough, and if $m\sim H(z_{eq})$, it is possible to generate a peak in the EDE fraction $f_{\rm ede}$ close to $z_{eq}$, at $z_{\rm max}$, that can lead to a decrease of the sound horizon and, therefore, force larger values of the Hubble rate at low redshifts. This is required to keep the position of the first peak of the CMB temperature angular power spectrum as it is measured by {\it Planck}, and also to keep the good fit to the BAO data. The decay of the EDE density is sufficiently fast to respect the tight constraints on the DE fraction around the recombination time \cite{Pettorino:2013ia,Gomez-Valent:2021cbe}. ULA was firstly proposed as a viable option to alleviate the Hubble tension in \cite{Poulin:2018cxd}. After that, some studies pointed out that this model tends to enhance the large-scale structure in the universe, see e.g. \cite{Hill:2020osr,DAmico:2020ods}. In order to mitigate the increase of the early integrated Sachs-Wolfe (iSW) effect introduced by EDE the model needs to increase $\omega_{\rm cdm}$ and $n_s$ as well, see e.g. \cite{Vagnozzi:2021gjh}. This, in turn, enhances the LSS if the other parameters of the theory (including $V_0$) do not allow for any compensation of this effect. There have been intensive discussions on this issue in the literature, see e.g. \cite{Hill:2020osr,DAmico:2020ods,Murgia:2020ryi,Smith:2020rxx,Gomez-Valent:2021cbe}, and \cite{Hill:2021yec,Poulin:2021bjr} for the results obtained incorporating also the data from the Atacama Cosmology Telescope \cite{ACT:2020gnv} and SPT-3G 2018 \cite{Smith:2022hwi}. These discussions have been mainly based on the interpretation of the posterior distributions obtained in several fitting analyses for the quantities $\sigma_8$ and $S_8=\sigma_8(\Omega^{(0)}_m/0.3)^{0.5}$, which tend to peak at larger values than those found for the $\Lambda$CDM and, therefore, to worsen the $\sigma_8/S_8$ tension. As firstly argued in \cite{Sanchez:2020vvb}, these quantities are sometimes difficult to interpret as clean LSS estimators, though. The reason is simple. The {\it rms} of mass fluctuations at the scale $R$ reads,

\begin{equation}\label{eq:rms}
\sigma_R^2=\frac{1}{2\pi^2}\int_0^{\infty}dk\,k^2P(k)[W(kR)]^2\,,
\end{equation}
with $R$ the radius of the spherical top-hat window function $W(kR)$ used to smooth the matter density field. The quantity $\sigma_8$ is computed using $R_8=8h^{-1}$ Mpc, which depends on the reduced Hubble parameter $h=H_0/(100\,{\rm km/s/Mpc})$. This means that in each step of a Monte Carlo analysis the value of $\sigma_8$ is not obtained using the same scale, just because the value of $h$ changes and so does also $R_8$. Thus, the fact that the posterior distribution of $\sigma_8$ is shifted towards larger values in a concrete model with respect to another model does not necessarily imply that matter fluctuations are larger in the former. If the posterior distribution for $H_0$ also peaks at higher values we systematically evaluate the {\it rms} of mass fluctuations at lower scales and this can produce an enhancement of $\sigma_8$ even if the shape of the matter power spectrum is the same or very similar in both models. 

The author of \cite{Sanchez:2020vvb} suggested to use the {\it rms} of mass fluctuations in spheres of radius $R_{12}=12$ Mpc, $\sigma_{12}$, and the related parameter $S_{12}=\sigma_{12}(\omega_{m}/0.14)^{0.4}$, instead of $\sigma_8$ and $S_8$. $R_{12}$ does not depend on $h$ and, hence, $\sigma_{12}$ is free from the aforementioned bias that might affect some models with posteriors for $h$ that differ from those found in the $\Lambda$CDM. Fitting values of these parameters were already obtained for the first time in \cite{Gomez-Valent:2021cbe} from the reconstruction of the EDE fraction with various data sets, showing that the values of $\sigma_{12}/S_{12}$ are in all cases lower than $\sigma_8/S_8$ and that the former lie closer to the $\Lambda$CDM results, which means that these matters can have a direct impact on the quantification of the tension between the models and the LSS data.

\begin{figure*}[t!]
\begin{center}
\includegraphics[width=5.3in,height=3.7in]{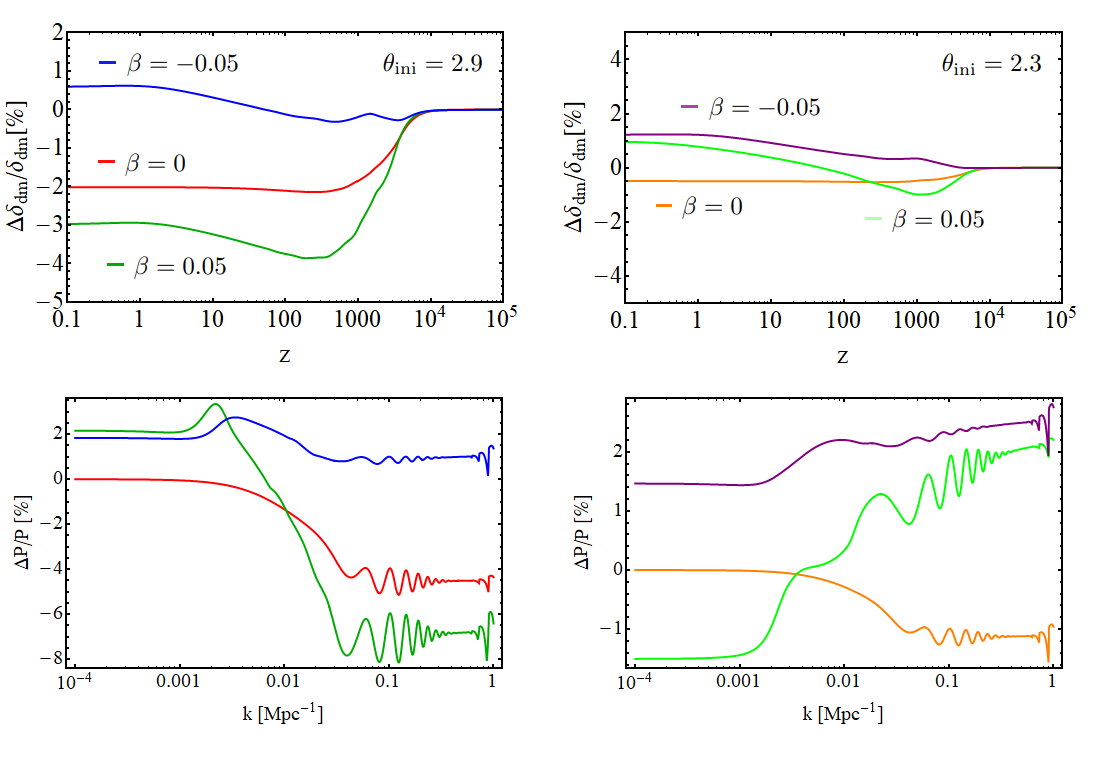}
\caption{{\it Upper plots:}  Relative differences between the density contrast of DM computed in the CDE\_ULA models and the $\Lambda$CDM (our reference model), as a function of the redshift and for $k=0.1$ Mpc$^{-1}$. For the standard model we use the values of the parameters obtained from the Planck 2018 TTTEEEE analysis \cite{Planck:2018vyg}, and for the CDE\_ULA models we use the same values of the parameters
as in Fig. \ref{fig:fig6}; {\it Lower plots:} Relative differences of the linear matter power spectrum at $z=0$, using the same models. }\label{fig:fig7}
\end{center}
\end{figure*}

In \cite{Gomez-Valent:2022hkb} it was shown that ULA actually alleviates the $H_0$ tension while keeping the value of $\sigma_{12}$ extremely close to the ones obtained in the standard model. This might be pointing out that ULA does not worsen the tension with the LSS data that much, if  at all. In this paper we will elaborate more on this observation, and will show explicitly that the best-fit ULA model does not suffer from any enhancement of the parameter $\sigma_{12}$. The equation that governs the evolution of matter fluctuations at $k\gg\mathcal{H}$ from the MDE onwards is the same as in the $\Lambda$CDM, i.e.

\begin{equation}\label{eq:dcEDE}
\frac{d^2\delta_m}{da^2}+\frac{3}{2a}\frac{d\delta_m}{da}[2-\Omega_{m}(a)]-\frac{3}{2a^2}\Omega_m(a)\delta_m(a)=0\,.
\end{equation}  
During the MDE $\Omega_m\approx 1$, so $\delta_m\sim a$ both in ULA and the concordance model. Larger values of $n_s$ make matter fluctuations to cluster more efficiently at large $k's$, though (at $k\gtrsim 0.1$ Mpc$^{-1}$). Nevertheless, ULA is able to accommodate values of the cosmological constant much higher than in the $\Lambda$CDM, making $\Omega_m$ to remain close to 0.3 at low redshifts. This helps to decrease the amplitude of $P(k)$ at larger scales (lower $k$'s). $P(k)$ is therefore larger than the one in the $\Lambda$CDM at small scales, but lower at large scales for typical best-fit values (cf. Sec. \ref{sec:results} and Fig. \ref{fig:fig11}). We will see later that there is a compensation in the computation of $\sigma_{12}$ for ULA with formula \eqref{eq:rms} (which involves an integral over a wide range of $k$'s) that renders its value stable. This might be pointing out that the assessment of the tension with the LSS data is more subtle than previously thought, and that it cannot be certainly carried out by just looking at the posteriors for $\sigma_8/S_8$. However, as we have mentioned before, despite the similarity in the values of $\sigma_{12}$ in ULA and the $\Lambda$CDM, there exist important differences in the shape of the matter power spectra. Probing observationally the clustering at different scales and measuring unambiguous LSS estimators (see e.g. \cite{Semenaite:2021aen}) can be a good way of arbitrating this issue. We will come back to this discussion in Sec. \ref{sec:results}. 
\newline
\newline
{\it Coupled dark energy with an ULA potential:} We call the ULA model with a non-null coupling (i.e. with $\beta\ne 0$) ``CDE\_ULA''. We suggested in the conclusions of our previous work \cite{Gomez-Valent:2021cbe} that a coupling in the dark sector could help, in principle, some EDE models with a non-negligible fraction of DE in some windows of the pre-recombination epoch to suppress the growth of matter fluctuations, while keeping large values of $H_0$. Some authors have already worked in this direction \cite{Karwal:2021vpk,McDonough:2021pdg}. In this paper we also explore this possibility for coupled ULA, trying to understand more in detail the phenomenology of the model and taking special care in the interpretation of the fitting results by providing constraints on some parameters that can help us to better assess the tensions, and also by studying the impact of volume effects in the marginalization process, which have been already found to be relevant in the uncoupled ULA model \cite{Herold:2021ksg,Gomez-Valent:2022hkb}.

Let us begin analyzing how the coupling affects the evolution of the EDE fraction and the dark matter mass in CDE\_ULA. If $z_{\max}\sim z_{\rm eq}$, the scalar field feels first the coupling, rather than the curvature of the potential. This is why positive values of $\beta$ allow to climb the latter up before rolling it down. This has two effects: (i) the moment at which $m\sim H$ happens at later times (i.e. lower redshifts) due to the increase of $\bar{\rho}_{\rm \phi}$; and (ii) $f_{\rm ede}$ is larger.  This can be seen in the two upper plots of Fig. \ref{fig:fig6}, for two different values of $\theta_{\rm ini}$. If $z_{\max}\gtrsim z_{\rm eq}$ these effects are also observed, but are less prominent and eventually become negligible when $z_{\max}\gg z_{\rm eq}$, just because the coupling does not have time to act alone, without the influence of the potential. 

Deep in the RDE the scalar field is in good approximation frozen and the DM mass remains almost constant, but the coupling injects some dynamics before $z_{\rm max}$ if $z_{\rm max}\sim z_{\rm eq}$. In this phase of the cosmic expansion if $\beta>0$ the scalar field grows mildly, whereas if the coupling is negative $\phi$ decreases. When the scalar field dynamics starts to be dominated by the potential, close to $z_{\rm max}$, we find that negative $\beta$'s produce a decrease of $m_{\rm dm}$ due to Eq. \eqref{eq:gmass} and the fact that $d\phi/da<0$. If $\beta>0$, instead, the DM mass can increase or decrease depending on the value of the scalar field at $z_{\rm max}$. If $\phi(z_{\rm max})>\pi f$ the DM mass decreases because $d\phi/da>0$, see again Eq. \eqref{eq:gmass}. If $\phi(z_{\rm max})<\pi f$ the DM mass increases because the opposite happens. This is why the initial condition of the scalar field plays also a role, as it is clear from the lower plots in Fig. \ref{fig:fig6}. If $\theta_{\rm ini}$ is close enough to $\pi$ and $\beta>0$ the scalar field can oscillate around $\phi=2\pi f$ after the decay instead of around $\phi=0$. If $z_{\rm max}\gg z_{\rm eq}$ this possibility is much less probable and requires either a fine tuning of $\theta_{\rm ini}$ to $\pi$ or larger (positive) values of the coupling to happen. Notice that the largest fraction of the change of $m_{\rm dm}$ happens right after $z_{\max}$, during the rolling-down period. The mean value of the DM mass is kept constant in the oscillatory phase.

\begin{table*}[t!]
\centering
\begin{tabular}{|c ||c | c || c| c|    }
 \multicolumn{1}{c}{} & \multicolumn{1}{c}{} & \multicolumn{1}{c}{} & \multicolumn{1}{c}{} & \multicolumn{1}{c}{} \\
\multicolumn{1}{c}{} &  \multicolumn{2}{c}{Planck18} &  \multicolumn{2}{c}{Planck18+SNIa+BAO}
\\\hline
{\small Parameter} & {\small $\Lambda$CDM}  & {\small CDE\_const}& {\small$\Lambda$CDM} & {\small CDE\_const}
\\\hline
$10^2\omega_b$ &  $2.239\pm 0.015$ (2.240) & $2.235^{+0.014}_{-0.015}$ (2.229) & $2.247^{+0.013}_{-0.014}$ (2.252) & $2.239^{+0.014}_{-0.015}$ (2.242) \\\hline
$\omega_{\rm cdm}$ & $0.1203^{+0.0011}_{-0.0013}$ (0.1209) & $0.1187^{+0.0028}_{-0.0012}$ (0.1194) & $0.1190\pm 0.0008$ (0.1189) & $0.1187\pm 0.0008$ (0.1180) \\\hline
$n_s$ & $0.966\pm 0.004$ (0.966) & $0.965^{+0.004}_{-0.005}$ (0.961) & $0.968\pm 0.004$ (0.966) & $0.966\pm 0.004$ (0.964) \\\hline
$\tau_{\rm reio}$ & $0.056^{+0.007}_{=0.006}$ (0.054) & $0.055^{+0.007}_{-0.008}$ (0.055) & $0.058^{+0.007}_{-0.008}$ (0.061) & $0.057^{+0.007}_{-0.008}$ (0.054) \\\hline
$\sigma_{12}$ & $0.807 \pm 0.008$ (0.809) & $0.811^{+0.009}_{-0.010}$ (0.819) & $0.799\pm 0.007$ (0.798)& $0.808^{+0.009}_{-0.010}$ (0.808) \\\hline
$H_0$ & $67.47^{+0.56}_{-0.48}$ (67.29) & $68.55^{+0.56}_{-1.73}$ (68.31) & $68.03^{+0.35}_{-0.36}$ (68.15) & $68.43^{+0.43}_{-0.53}$ (68.95)\\\hline
$\beta$ & $-$ & $<0.040$ (0.040) & $-$ & $0.028^{+0.017}_{-0.013}$ (0.041) \\\hline\hline
$r_d$ & $146.93^{+0.25}_{-0.27}$ (146.75) & $146.88^{+0.26}_{-0.28}$ (146.69) & $147.18^{+0.20}_{-0.22}$ (147.15)& $146.99^{+0.27}_{-0.24}$ (146.89) \\\hline
$M$ & $-$ & $-$ & $-19.408\pm 0.010$ (-19.404) & $-19.397^{+0.012}_{-0.015}$ (-19.381) \\\hline
$S_{8}$ & $0.832^{+0.012}_{-0.013}$ (0.836) & $0.827^{+0.015}_{-0.013}$ (0.838) & $0.818^{+0.010}_{-0.009}$ (0.816)& $0.824^{+0.010}_{-0.011}$ (0.821) \\\hline
$\sigma_8$ & $0.814\pm 0.005$ (0.814) & $0.827^{+0.007}_{-0.021}$ (0.833) & $0.811\pm 0.006$ (0.810)&  $0.822^{+0.009}_{-0.013}$ (0.827)
 \\\hline
$S_{12}$ & $0.813^{+0.010}_{-0.009}$ (0.817)  & $0.814\pm 0.010$ (0.823) & $0.803\pm 0.008$ (0.801)  & $0.810^{+0.009}_{-0.010}$ (0.809)  \\\hline
$\Delta\chi^2_{\rm min}$ & $-$  & 1.93   & $-$ & 2.39
\\\hline
$\Delta{\rm AIC}$ & $-$  & -0.07   & $-$ & 0.39
\\\hline
\end{tabular}
\caption{Mean and $1\sigma$ uncertainties of the individual parameters of the $\Lambda$CDM and CDE\_const models, together with the corresponding best-fit values (between parenthesis). $H_0$ and $r_d$ are expressed in km/s/Mpc and Mpc, respectively. In the last two rows we report the differences $\Delta\chi^2_{\rm min}\equiv \chi^2_{\Lambda,{\rm min}}-\chi^2_{{\rm CDE\_const},{\rm min}}$ and $\Delta {\rm AIC}\equiv{\rm AIC}_{\Lambda}-{\rm AIC}_{\rm CDE\_const}$ to better assess the fitting performance of the aforementioned models. We discuss these results in Sec. \ref{sec:results}. }
\label{tab:tableCDE}
\end{table*}

In the upper plots of Fig. \ref{fig:fig7} we show how the wave mode $k=0.1$ Mpc$^{-1}$ of the DM density contrast evolves in the ULA and CDE\_ULA models with respect to the $\Lambda$CDM, for fixed initial conditions and two alternative values of $\theta_{\rm ini}$. As expected, the larger is $f_{\rm ede}$ the larger is the relative decrease of $\delta_{\rm dm}$ at $\sim z_{\rm max}$, since dark energy fights against the aggregation of matter. After the decay of the scalar field, in ULA the relative difference with respect to the $\Lambda$CDM result remains constant, since the equation for the density contrast \eqref{eq:dcEDE} is the same to the one in the standard model. In CDE\_ULA, though, the fifth force makes $\delta_{\rm dm}$ to grow faster in the MDE. This growth can compensate in most of the cases the negative bump generated around $z_{\rm max}$, giving rise to an increase of the DM density contrast at low redshifts, but there are also cases in which this compensation is not complete (see e.g. the green curve in the left upper plot of Fig. \ref{fig:fig7}). The oscillations of the scalar field deep in the MDE (see the inner plots in Fig. \ref{fig:fig6}) make the friction term of Eq. \eqref{eq:densityContrastDM} to be inefficient, so in CDE\_ULA the equation of the DM density constrast reads in good approximation,

\begin{equation}
\delta_{\rm dm}^{\prime\prime}+\mathcal{H}\delta^\prime_{\rm dm}-\frac{\kappa^2a^2}{2}\left[\bar{\rho}_b\delta_b+\bar{\rho}_{\rm dm}\delta_{\rm dm}(1+2\beta^2)\right]=0\,.
\end{equation}
During the scaling regime in the MDE, when the matter fraction oscillates around $\Omega_{m}\sim 1-2\beta^2/3$, $\delta_{\rm dm}\sim a^{1+6\beta^2/5}$ \cite{McDonough:2021pdg}, so matter fluctuations grow in this epoch slower than in CDE\_const, but faster than in the concordance model. The relative differences between the matter power spectrum of the CDE\_ULA models and the $\Lambda$CDM are shown in the lower plots of Fig. \ref{fig:fig7}. We remark that these plots have been obtained fixing the initial conditions. By changing them, and the parameters of the EDE scalar field potential we can make the density perturbations to evolve differently, of course. Here we only wanted to show what is the impact of the coupling. For a precise assessment of the ability of these models to describe the current cosmological data and loosen the cosmological tensions we are forced to perform a detailed fitting analysis. We devote the next sections to explain the fitting strategy (data sets and methodology) and report our results.


\begin{figure*}[t!]
\begin{center}
\includegraphics[width=6.3in,height=5.7in]{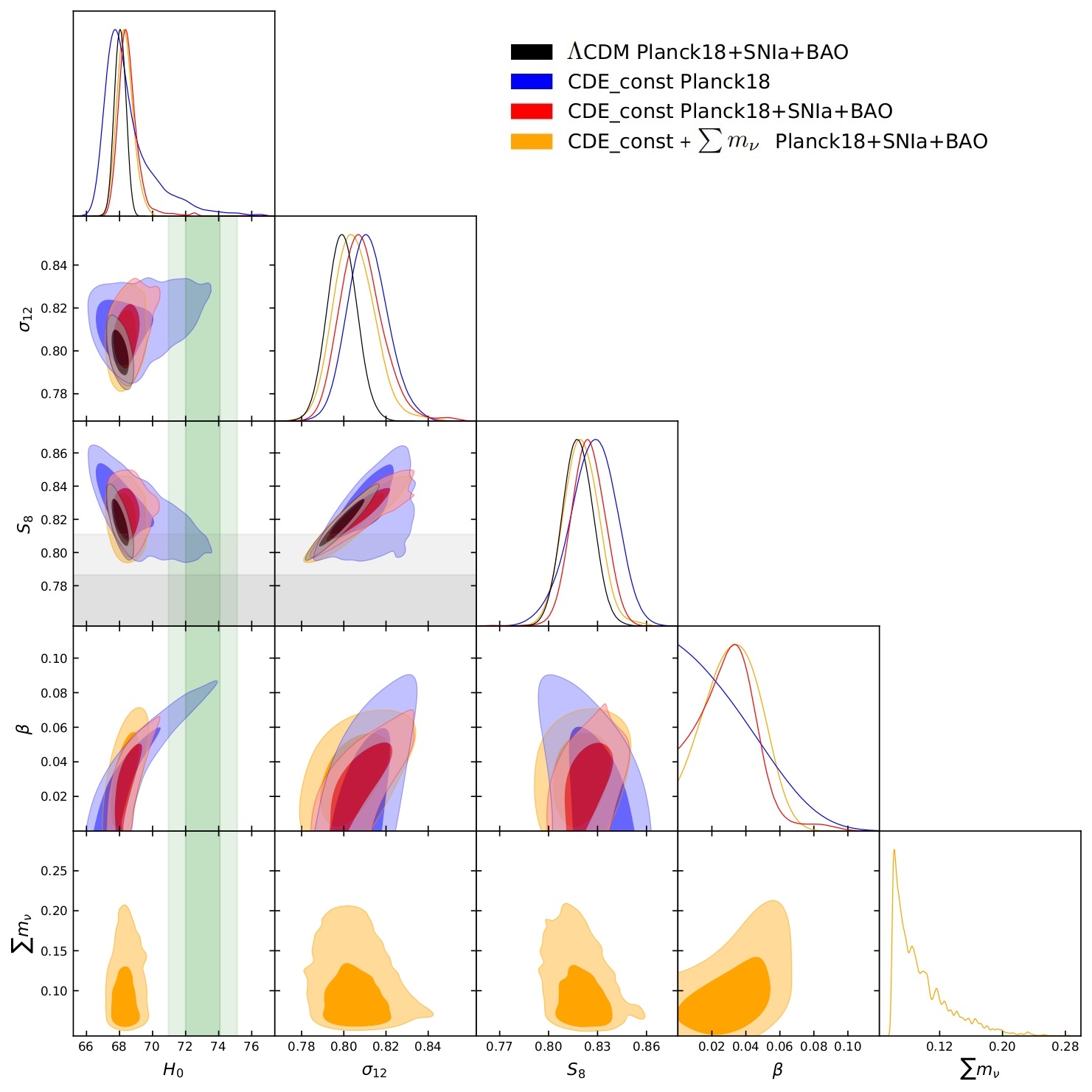}
\caption{One-dimensional posterior distributions for some parameters of the $\Lambda$CDM and CDE\_const, and the corresponding contour plots at $1\sigma$ and $2\sigma$ c.l. The constraints from neutrino experiments (cf. Sec. \ref{sec:data}) do not allow  $\sum m_\nu$ to be lower than $\sim 0.06$ eV. The green vertical bands in the first column correspond to the measurement of $H_0$ by the SH0ES team \cite{Riess:2021jrx}, whereas the gray horizontal bands are the weak lensing constrain on $S_8$ from KiDS+VIKING-450+DES-Y1, which is obtained under the assumption of the $\Lambda$CDM \cite{Joudaki:2019pmv}. Both are provided at $1\sigma$ and $2\sigma$ c.l. See the comments in Sec. \ref{sec:results}.}\label{fig:fig8}
\end{center}
\end{figure*}

\section{Data sets}\label{sec:data}

\noindent These are the data sets employed in our fitting analyses:
\newline
\newline
\noindent {\it CMB data:} We employ the full {\it Planck} 2018 TTTEEE+lowE+lensing likelihood \cite{Planck:2018vyg}, i.e. the data on the temperature (TT) and polarization (EE) anisotropies of the cosmic microwave background (CMB), their cross-correlations (TE), and the data on the CMB lensing reconstruction. We vary in our Monte Carlo (MC) runs the 21 {\it Planck} nuisance parameters together with the cosmological ones. We denote this data set as Planck18, in short.
\newline
\newline
{\it Supernovae and BAO:} We also fit our models to a richer data set that incorporates the observational ingredients usually used to build the cosmic inverse distance ladder \cite{Aubourg:2014yra,Cuesta:2014asa,Feeney:2018mkj,Camarena:2019rmj}, which is relevant for the discussion on the $H_0$ tension. We combine Planck18 with the data on SNIa from the Pantheon compilation \cite{Scolnic:2017caz} and the BAO information from several galaxy surveys. We call this data set Planck18+SNIa+BAO. The absolute magnitude of the SNIa, $M$, is left free in the MC runs. We display its value in our tables in all cases in which we employ SNIa. It is a relevant quantity. The SH0ES collaboration has measured $M$ using the calibration of the SNIa in the first steps of the (direct) cosmic distance ladder, $M_{\rm SH0ES}=-19.253\pm 0.027$ \cite{Riess:2021jrx}, which is independent from the underlying cosmology. This value is fully compatible with the one inferred in model-independent studies of low-redshift cosmological data sets \cite{Gomez-Valent:2021hda,Benisty:2022psx}, but much higher than the one obtained in cosmological analyses of the $\Lambda$CDM using the CMB data from {\it Planck} and high-redshift SNIa. The latter usually lies close to $M\sim -19.40$. The $H_0$ tension can be thought of as a tension between the locally measured value of $M$ and the one inferred from cosmological studies in the context of the standard model, see e.g. \cite{Camarena:2019moy,Efstathiou:2021ocp}. It can be therefore enlightening to discuss the $H_0$ tension directly in terms of $M$. We do so in Sec. \ref{sec:results}.

\begin{table*}[t!]
\centering
\begin{tabular}{|c ||c | c || c| c|    }
 \multicolumn{1}{c}{} & \multicolumn{1}{c}{} & \multicolumn{1}{c}{} & \multicolumn{1}{c}{} & \multicolumn{1}{c}{} \\
\multicolumn{1}{c}{} &  \multicolumn{2}{c}{Planck18} &  \multicolumn{2}{c}{Planck18+SNIa+BAO}
\\\hline
{\small Parameter} & {\small EXP}  & {\small CDE\_EXP}& {EXP} & {\small CDE\_EXP}
\\\hline
$10^2\omega_b$ &  $2.236\pm 0.016$ (2.235) & $2.256^{+0.020}_{-0.019}$ (2.247)  &  $2.248^{+0.014}_{-0.013}$ (2.250)   &  $2.247^{+0.016}_{-0.026}$ (2.256) \\\hline
$\omega_{\rm cdm}$ & $0.1210^{+0.0015}_{-0.0015}$ (0.1201) & $0.1229^{+0.0023}_{-0.0030}$ (0.1211) & $0.1191^{+0.0008}_{-0.0009}$ (0.1192)   & $0.1189^{+0.0009}_{-0.0009}$ (0.1195)  \\\hline
$n_s$ & $0.964\pm 0.004$ (0.965) & $0.964\pm 0.005$ (0.965) & $0.968\pm 0.004$ (0.969)  & $0.967 \pm 0.004$ (0.968)  \\\hline
$\tau_{\rm reio}$ & $0.056^{+0.007}_{-0.008}$ (0.056) & $0.051\pm 0.008$ (0.042) & $0.060^{+0.007}_{-0.008}$ (0.058)  & $0.055^{+0.010}_{-0.007}$ (0.051) \\\hline
$\sigma_{12}$ & $0.805^{+0.009}_{-0.008}$ (0.803) & $0.814^{+0.011}_{-0.015}$ (0.821) & $0.797\pm 0.007$ (0.802) & $0.811^{+0.010}_{-0.015}$ (0.824) \\\hline
$H_0$ & $67.20^{+0.64}_{-0.63}$ (67.45) & $65.43^{+1.95}_{-1.48}$ (66.54) & $68.01\pm 0.37$ (67.97) & $68.28^{+0.58}_{-0.70}$ (67.72)  \\\hline
$\Omega_{\rm ede}^{\rm RD}$ [\%] & $<0.22$ (0.06) & $<0.29$ (0.03) & $<0.13$ (0.00)  & $<0.13$ (0.00)  \\\hline
$\beta$ & $-$ & $-0.061^{+0.018}_{-0.038}$ (-0.079) & $-$ & $-0.008^{+0.090}_{-0.115}$ (-0.055) \\\hline\hline
$r_d$ & $146.65^{+0.42}_{-0.28}$ (146.97)  & $146.83^{+0.41}_{-0.35}$ (147.09) & $147.07^{+0.24}_{-0.22}$ (147.10)  & $146.99^{+0.36}_{-0.48}$ (147.12) \\\hline
$M$ & $-$ & $-$ & $-19.409\pm 0.010$ (-19.405)  & $-19.401^{+0.016}_{-0.019}$ (-19.417) \\\hline
$S_{8}$ & $0.833\pm  0.014$ (0.827) & $0.855\pm 0.021$ (0.852) & $0.817\pm 0.010$ (0.821) & $0.829^{+0.011}_{-0.017}$ (0.846)\\\hline
$\sigma_8$ & $0.810^{+0.008}_{-0.006}$ (0.809) & $0.803^{+0.019}_{-0.021}$ (0.820) & $0.808\pm 0.006$ (0.812)  & $0.825^{+0.009}_{-0.015}$ (0.833)
 \\\hline
$S_{12}$ & $0.813\pm 0.011$ (0.809)  & $0.826^{+0.014}_{-0.016}$ (0.829) & $0.801\pm 0.008$ (0.805) & $0.814^{+0.010}_{-0.014}$ (0.829)  \\\hline
$\Delta\chi^2_{\rm min}$ & 2.75  & 7.51  & -0.15 & 3.75
\\\hline
$\Delta{\rm AIC}$ & 0.75  & 3.51   & -2.15 & -0.25
\\\hline
\end{tabular}
\caption{As in Table \ref{tab:tableCDE}, but for the EXP and CDE\_EXP models.}
\label{tab:tableEXP}
\end{table*}

Our BAO data set includes the following points:

\begin{itemize}

\item $D_V/r_d$ at $z=0.122$ provided in \cite{Carter:2018vce}, with $D_V$ being the dilation scale,

\begin{equation}
D_V(z)=\left[D_M^2(z)\frac{cz}{H(z)}\right]^{1/3}\,,
\end{equation}
and $D_M=(1+z)D_{A}(z)$ the comoving angular diameter distance. This data point combines the dilation scales previously reported by the 6dF Galaxy Survey (6dFGS) \cite{Beutler:2011hx} at $z=0.106$ and the Sloan Digital Sky Survey (SDSS) Main Galaxy Sample at $z=0.15$ \cite{Ross:2014qpa}. 

\item The anisotropic BAO data ($D_A(z)/r_d$, $H(z)r_d$) measured by BOSS using the LOWZ ($z=0.32$) and CMASS ($z=0.57$) galaxy samples \cite{Gil-Marin:2016wya}.

\item The dilation scale measurements by WiggleZ at $z=0.44,0.60,0.73$ \cite{Kazin:2014qga}.

\item $D_A(z)/r_d$ at $z=0.81$ measured by the Dark Energy Survey (DES) \cite{DES:2017rfo}.

\item  The anisotropic BAO data from the extended BOSS Data Release 16 (DR16) quasar sample at $z=1.48$ \cite{Neveux:2020voa}.

\item The anisotropic BAO information obtained from the Ly$\alpha$ absorption and quasars of the final data release (SDSS DR16) of eBOSS, at $z=2.334$ \cite{duMasdesBourboux:2020pck}.

\end{itemize}
{\it A comment on BBN:} The only models among those described in Sec. \ref{sec:phenome} that could modify significantly the big bang nucleosynthesis (BBN) processes are those with the exponential potential discussed in Sec. \ref{sec:expPot}, i.e. EXP and CDE\_EXP, since we assume that the scaling solution in these models is already reached at $z_{\rm ini}=10^{14}\gg z_{\rm BBN}$. Nevertheless, we will see in Sec. \ref{sec:results} that we get very tight constraints on $\Omega_{\rm ede}^{\rm RD}$ from the CMB data, i.e. using CMB data alone we obtain a way lower upper bounds than those imposed by the BBN data \cite{Uzan:2010pm}, so even in these models EDE does not have a sizable impact on the physics at the BBN epoch. On the other hand, the upper value of the flat prior employed for $z_{\rm max}$ in ULA and CDE\_ULA (cf. Sec. \ref{sec:ULAPot}), is five orders of magnitude lower than $z_{\rm BBN}$. Hence, the EDE fraction is completely negligible at the BBN time and therefore does not affect the nucleosynthesis physics either. Some effects can be introduced by shifts of the baryon energy density in this model, but they are in any case very small \cite{Seto:2021xua}. In summary, we do not need to consider any BBN constraint in our study. 
\newline
\newline
{\it Massive neutrinos:} In most of our fitting analyses we use two massless and one massive neutrino of $0.06$ eV. However, we also constrain some models using the Planck18+SNIa+BAO data set allowing the three neutrino masses to vary simultaneously in the Monte Carlo runs, assuming a normal ordering, i.e. $m_1<m_2<m_3$. The latter is a sufficient assumption because the constraints on the sum of the neutrino masses $\sum m_\nu$ are not very sensitive to the neutrino hierarchy   \cite{Loureiro:2018pdz}. We consider the constraints from solar and atmospheric neutrinos, which read, respectively \cite{Esteban:2016qun}: $\Delta m_{21}^2\equiv m_2^2-m_1^2=(7.50^{+0.19}_{-0.17})\times 10^{-5}\,{\rm eV^2}$ and $\Delta m_{31}^2\equiv |m_3^2-m_1^2|=(2.524^{+0.039}_{-0.040})\times 10^{-3}\,{\rm eV^2}$. We actually vary $m_1$, $\Delta m^2_{21}$ and $\Delta m^2_{31}$ in the Monte Carlo, and compute the masses of the two heavier neutrinos using 

\begin{figure*}[t!]
\begin{center}
\includegraphics[width=5.3in,height=5.0in]{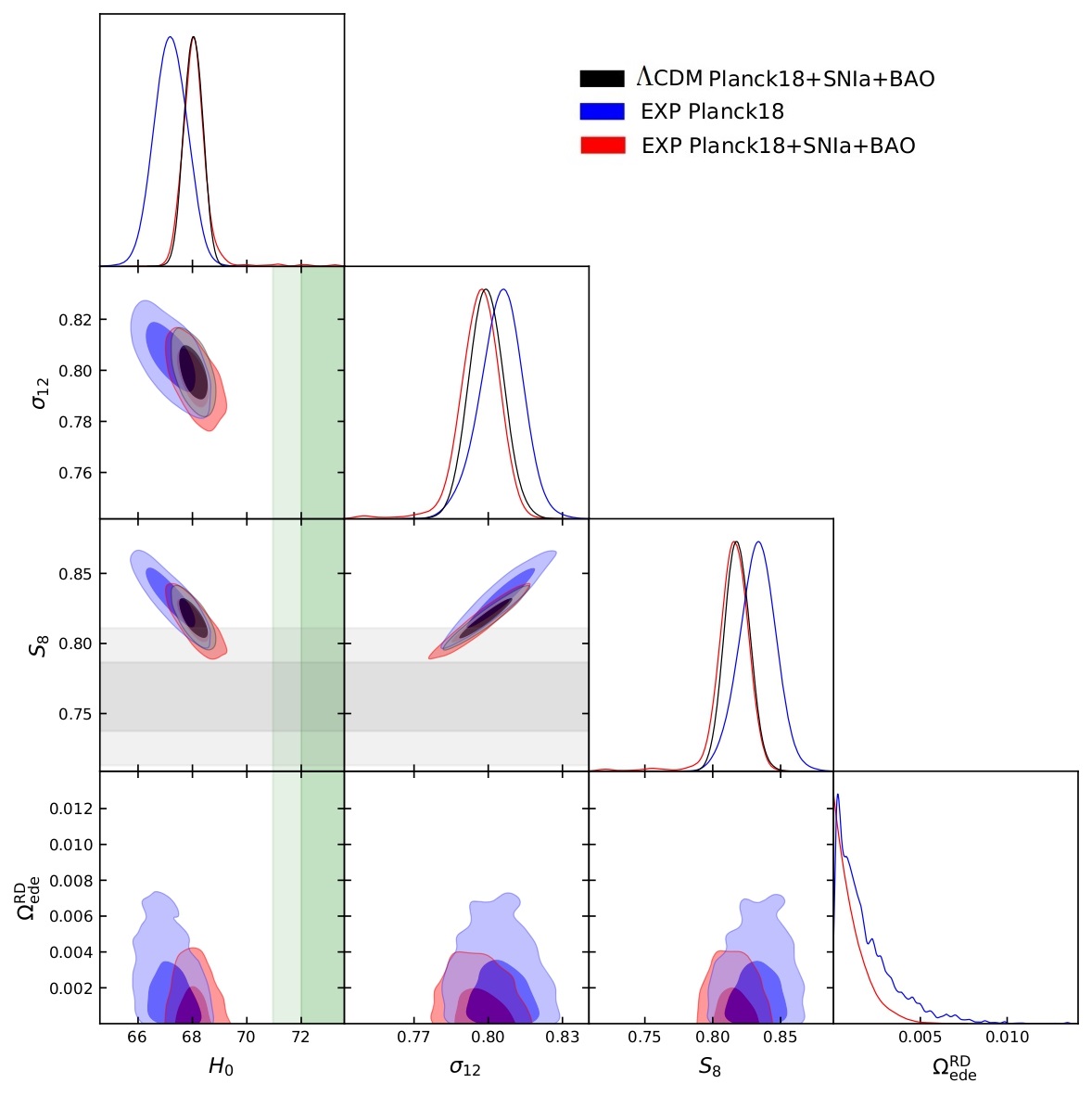}
\caption{Triangle plot for the EXP model, with the parameters that are relevant for the discussion of the cosmological tensions. As in Fig. \ref{fig:fig8} and the other triangle plots in this paper, we show the constraint from SH0ES \cite{Riess:2021jrx} in green and the one from KiDS+VIKING-450+DES-Y1 \cite{Joudaki:2019pmv} in gray.}\label{fig:fig9}
\end{center}
\end{figure*}

\begin{equation}
 m_2=\sqrt{m_1^2+|\Delta m^2_{21}|}\quad ;\quad m_3=\sqrt{m_1^2+|\Delta m^2_{31}|}   
\end{equation}
Massive neutrinos reduce the clustering of matter below the free streaming scale (see e.g. \cite{Lesgourgues:2006nd}) and, therefore, could have an impact on our analyses if the data give room to larger values of $\sum m_\nu$. \newline
\newline
We display the fitting results for the models discussed in Sec. \ref{sec:phenome} under the Planck18 and Planck18+SNIa+BAO data sets in Sec. \ref{sec:results}.

\begin{figure*}[t!]
\begin{center}
\includegraphics[width=6.1in,height=5.8in]{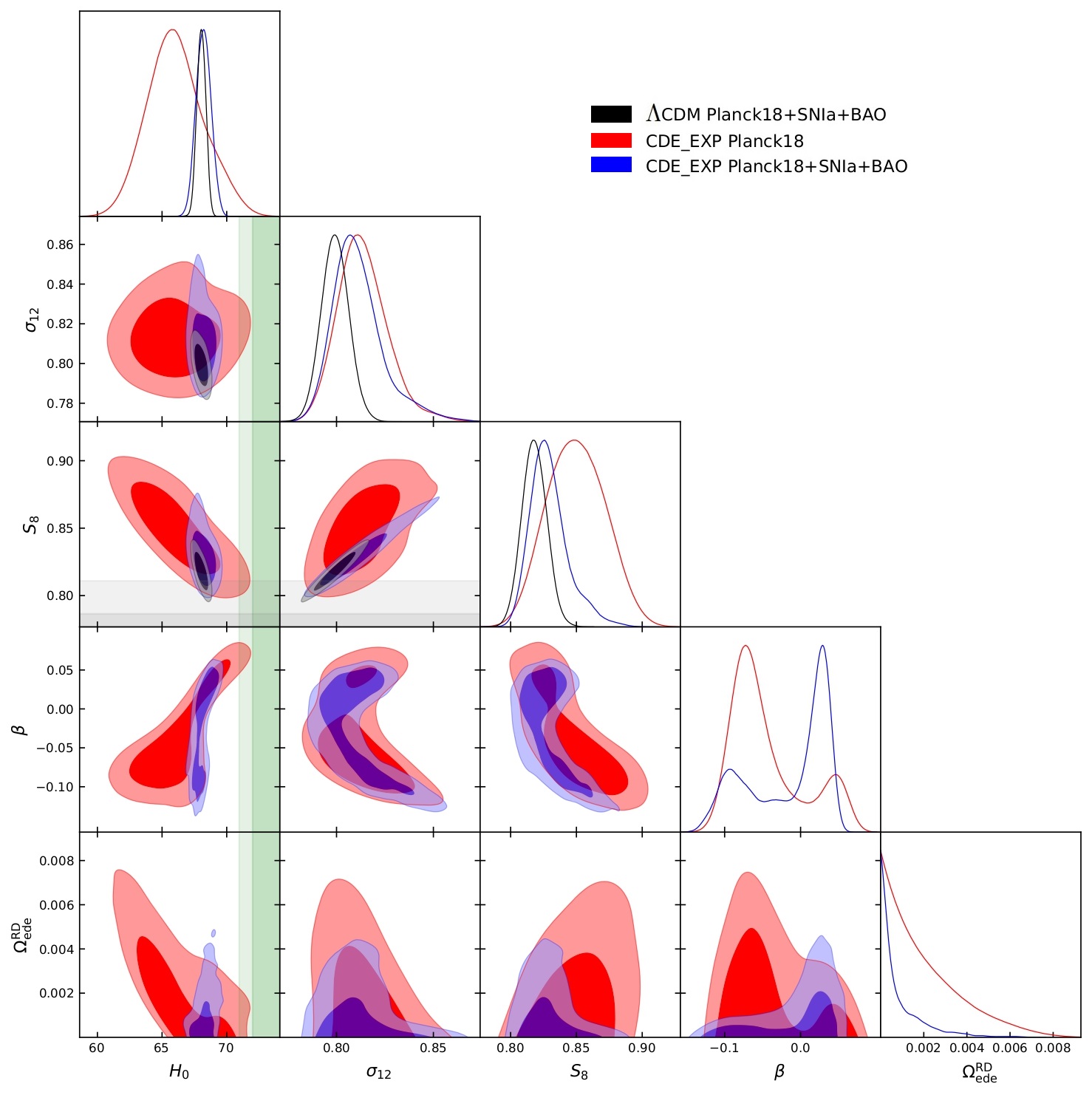}
\caption{Contour plots at $1\sigma$ and $2\sigma$ c.l. and one-dimensional posterior distributions for the CDE\_EXP model, obtained with the Planck18 and Planck18+SNIa+BAO data sets.}\label{fig:fig9b}
\end{center}
\end{figure*}


\section{Method}\label{sec:method}

We have implemented the background and linear perturbations equations of Sec. \ref{sec:eqs} and the various scalar field potentials presented and discussed in Sec. \ref{sec:phenome} in our own modified versions of the Einstein-Boltzmann system solver \texttt{CLASS} \cite{Blas:2011rf}. All of them have gone, of course, through the corresponding validation processes in order to check their correct functioning. For instance, we have checked that we recover the expected scaling solutions in CDE\_EXP and their nested models CDE\_const and EXP; and for ULA we have checked that we are able to reproduce other results found in the literature.

We have used \texttt{MontePython} \cite{Audren:2012wb} to sample the likelihood with the Metropolis-Hastings algorithm \cite{Metropolis:1953aaa,Hastings:1970bbb}. It is constructed from the data sets described in Sec. \ref{sec:data}. The Markov chains obtained in the output have been processed also with \texttt{MontePython} to obtain the marginalized constraints for the individual parameters and the absolute best-fit values, and with \texttt{GetDist} \cite{Lewis:2019xzd} to generate the confidence contour lines in all the relevant planes of parameter space and the marginalized one-dimensional posteriors. The minimum values of the $\chi^2$ in each run, i.e. $\chi^2_{\rm min}$, and the corresponding best-fit values of the parameters  have been found applying the procedure described in Appendix D.1 of \cite{Schoneberg:2021qvd}. We report our results in the tables and figures of the next section. 

For some parameters of CDE\_EXP, ULA and CDE\_ULA we have also applied the profile distribution method to get rid of the volume effects that could affect them, see e.g. \cite{Trotta:2017wnx}. We comment on these results in Sec. \ref{sec:results}. For previous analyses using this statistical technique see \cite{Planck:2013nga,Herold:2021ksg,Gomez-Valent:2022hkb,Campeti:2022vom}. We use the same approach explained in \cite{Gomez-Valent:2022hkb}, i.e. we compute the individual profile distributions directly from the Markov chains, instead of carrying out a series of minimization runs. This allows us to save a considerable amount of computing time.

In order to minimize the computational time spent in our Monte Carlo runs, we have tried to modify \texttt{CLASS} avoiding the use of the shooting method. This is impossible, though, in the case of ULA with and without coupling, since in this case we need to sample the distribution using the pair of parameters $(f_{\rm ede},z_{\rm max})$ instead of $(m,f)$, cf. Sec. \ref{sec:ULAPot}. This is more reasonable from the cosmological perspective, and not doing so could hide the evidence for a non-null EDE fraction $f_{\rm ede}$ due to prior issues \cite{Hill:2020osr}. We explain the procedure that we have followed to implement the shooting method for these models in Appendix B.


\section{Results and discussion}\label{sec:results}

We report our main fitting results in Tables \ref{tab:tableCDE}-\ref{tab:tableMassive} and Figs. \ref{fig:fig8}-\ref{fig:fig15}. In Appendix C we present some additional tables, where we list the contribution of the $\chi^2_i$'s of the individual data sets to the minimum value $\chi^2_{\rm min}$ in each model.

\begin{figure*}[t!]
\begin{center}
\includegraphics[width=6.1in,height=5.0in]{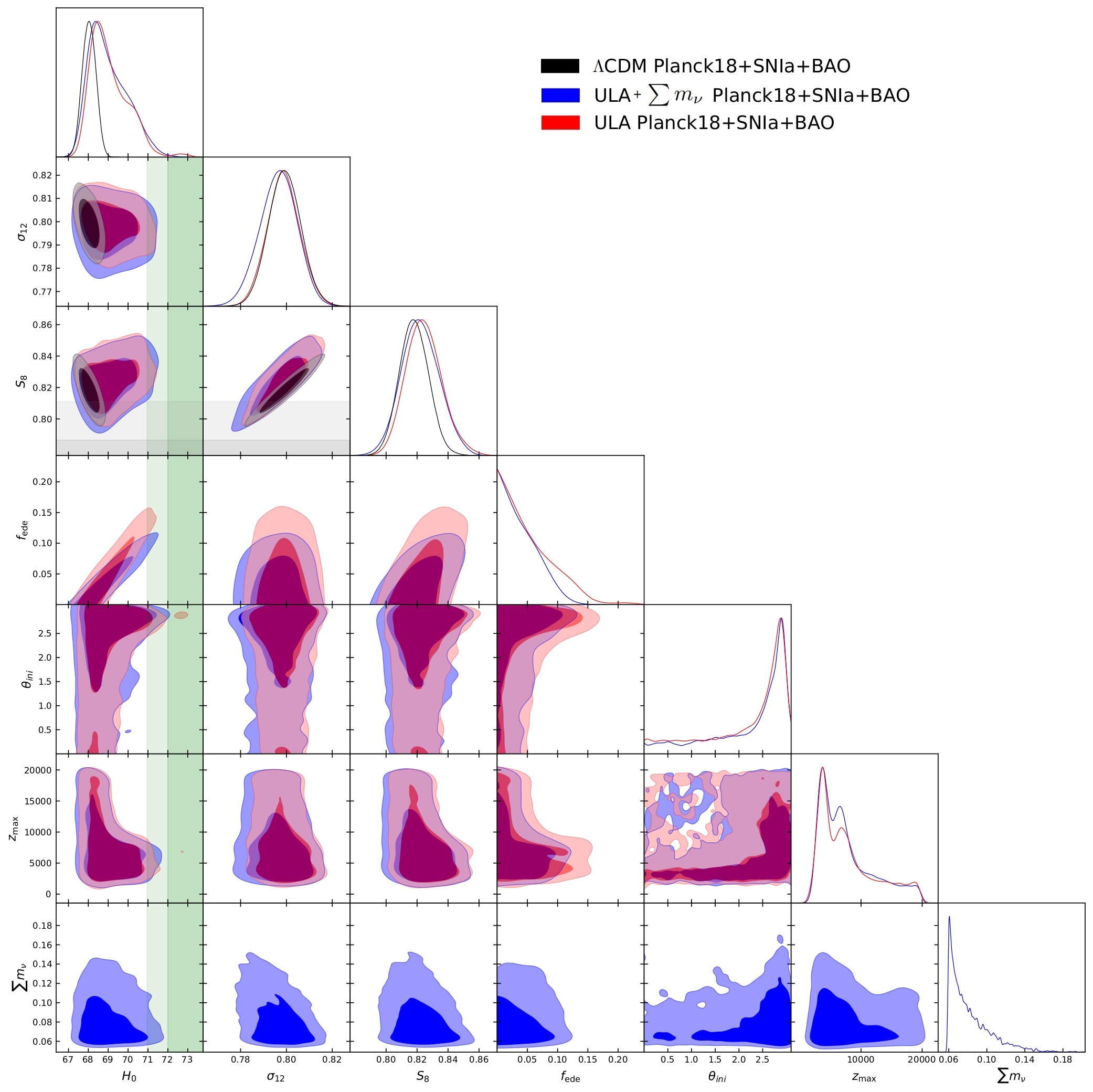}
\caption{Confidence regions at $1\sigma$ and $2\sigma$ c.l. and one-dimensional posterior distributions obtained with the Planck18+SNIa+BAO data set with the $\Lambda$CDM and ULA. The parameters $f_{\rm ede}$,  $z_{\rm max}$ and $\theta_{\rm ini}=\phi_{\rm ini}/f$ are the maximum EDE fraction, the redshift at which there is the peak and the initial value of the scalar field at $z_{\rm ini}=10^{14}$, respectively (cf. Sec. \ref{sec:phenome}). For ULA+$\sum m_\nu$ we allow the sum of the neutrino masses to vary in the Monte Carlo analysis, using the approach described in Sec. \ref{sec:data}.}\label{fig:fig10}
\end{center}
\end{figure*}

We discuss first the output of the analysis of the $\Lambda$CDM and CDE\_const. These two models are nested. The former is described in terms of six cosmological parameters, whereas the latter has an additional parameter, the coupling $\beta$. It controls the departure of CDE\_const from the $\Lambda$CDM. We find that the CMB data from Planck does not exclude (when used alone) values of $\beta\sim 0.07$ at $2\sigma$ c.l., but the value of the coupling is still fully compatible with 0, being $\beta<0.04$ at $1\sigma$ c.l. (see the second column of Table \ref{tab:tableCDE}). In Fig. \ref{fig:fig8} we observe a positive correlation between the coupling and the Hubble parameter, as expected \cite{Pettorino:2013ia,Planck:2015bue,Gomez-Valent:2020mqn}. Also between $\beta$ and the LSS estimator $\sigma_{12}$. The reasons have been already discussed in Sec. \ref{sec:phenome}. A larger coupling increases the growth of matter perturbations during the MDE and requires a larger expansion rate at low redshifts to not alter the location of the CMB peaks and the description of the BAO data. The contours in the $(H_0,\beta)$-plane show that it is possible to obtain values of $H_0$ compatible with the distance ladder measurement from SH0ES \cite{Riess:2021jrx} at $2\sigma$ c.l., due the long right tale of the distribution. There is, though, a shift towards larger values of $\sigma_{12}$, which nevertheless is not drastically big. When we add the SNIa+BAO data, we can see that the values of $\beta\gtrsim 0.06$ get strongly excluded. This, in turn, removes the tail of the one-dimensional posterior of $H_0$, limiting its value to $H_{0}=(68.43^{+0.43}_{-0.53})$ km/s/Mpc at $1\sigma$ c.l. and keeping the $H_0$ tension at more than $3\sigma$. Although the difference between $\Delta\chi^2=\chi^2_{\Lambda,{\rm min}}-\chi^2_{\rm CDE\_const,min}=2.39$ is small and certainly there is no significant statistical preference for an interaction in the dark sector, it is worth to point out the existence of a peak in the posterior of $\beta$ when we include the full data set. The maximum is located $\sim 2\sigma$ away from 0, and the mean reads $\beta=0.028^{+0.017}_{-0.013}$ at 68.2\% c.l. This peak has also been found in past analyses using similar data configurations, see \cite{Pettorino:2013ia,Planck:2015bue,Gomez-Valent:2020mqn}, and it was recently shown in \cite{Gomez-Valent:2022hkb} that it is not sourced by spurious marginalization effects. It allows a better description of the CMB data (cf. Table \ref{tab:tableB2}).

By allowing the mass of the neutrino masses to vary in the analysis we find very similar results, see Table \ref{tab:tableMassive}. The peak is still present, and slightly displaced to the right ($\beta=0.033\pm 0.015$) due to the positive correlation between the coupling and $\sum m_\nu$, which allows to compensate the enhancement of the matter power spectrum caused by the fifth-force. This positive correlation, which is more evident in the range $\beta>0.03$, also explains the weaker constraints on $\sum m_\nu$, which in CDE\_const is $<0.18$ eV at $95\%$ c.l., whereas in the $\Lambda$CDM $<0.12$ eV. Lower values of the parameter $\sigma_{12}$ are allowed due to the suppression of power at low scales introduced by the massive neutrinos, but the differences are small. The contours in the $(H_0,\beta)$-plane of Fig. \ref{fig:fig8} show, as expected, that the large values of $\beta$ are less able to lead to larger values of the Hubble parameter than in the case in which we consider a minimal neutrino setup with only one massive neutrino of 0.06 eV. 

We use the Akaike information criterion (AIC) \cite{Akaike} to penalize the additional degrees of freedom and perform a fairer comparison of the statistical performance of the models. For a sufficiently large number of data points, the AIC is defined as

\begin{equation}\label{eq:AIC}
{\rm AIC}= \chi^2_{\rm min}+2{\rm k},  
\end{equation}
with k the number of parameters of the model. The difference between the AIC values obtained in two models (1 and 2) can be used as an approximation of the logarithm of the Bayes ratio $B_{12}$, ${\rm AIC}_{12}={\rm AIC}_{2}-{\rm AIC}_{1}\approx 2\ln{B_{12}}$, with $B_{12}=\mathcal{E}_{2}/\mathcal{E}_1$ and $\mathcal{E}_i$ the Bayesian evidence of model $i$. The small differences $\Delta{\rm AIC}\lesssim \mathcal{O}(1)$ between the $\Lambda$CDM and CDE\_const confirm that from the Bayesian point of view the two models perform similarly well in the description of the Planck18 and Planck18+SNIa+BAO data sets. There is a slight decrease of the $\chi^2_{\rm min}$ in CDE\_const, but the addition of the coupling $\beta$ is not especially favored in the light of Occam's razor. 

EDE with an exponential potential \eqref{eq:expPot} is already very constrained by the CMB data, which limits $\Omega_{\rm ede}^{\rm RD}<0.22\%$ at 1$\sigma$ c.l., basically due to the tight upper bound that the EDE fraction has to satisfy at the decoupling time \cite{Gomez-Valent:2021cbe}. Our fitting results for this model are reported in Table \ref{tab:tableEXP}. There is no room in this model for an alleviation of the cosmological tensions. The addition of SNIa+BAO gives even less margin, since $\Omega_{\rm ede}^{\rm RD}<0.13\%$ at 1$\sigma$ c.l. In Fig. \ref{fig:fig9} we can see that the constraints on $\sigma_{12}$ and $H_0$ are actually very similar to those found in the $\Lambda$CDM, with a negative correlation between these two parameters and with a peak of the posterior lying very far away from the SH0ES bands. We have checked that volume effects are unimportant in this model.

CDE\_EXP allows us to explain much better than the $\Lambda$CDM the CMB data. The corresponding $\chi^2_{\rm min}$ is 7.51 units below the values obtained in the standard model, and this is thanks to a $2\sigma$ departure of the coupling from zero, with the posterior of $\beta$ peaking in the negative region. Notice that due to the non-trivial potential \eqref{eq:expPot}, the coupling can acquire larger values than in CDE\_const. The regions RGIIa and RGI of parameter space are preferred (see Figs. \ref{fig:fig1} and \ref{fig:fig9b}), being the former more probable when we allow for larger values of the EDE fraction in the RDE. This is supported also by the AIC. We obtain $\Delta{\rm AIC}=+3.51$, which according to Jeffreys' scale corresponds to a moderate evidence for CDE\_EXP. However, this hint for new physics is washed out by the SNIa and BAO data,  cf. Table \ref{tab:tableEXP}. In Fig. \ref{fig:fig9b} we cannot appreciate any decrease of $\sigma_{12}$ compared to the $\Lambda$CDM, although the model allows for larger values of $H_0$. The tension is only mildly loosened, though, at a similar level at which it is in CDE\_const. Under the Planck18+SNIa+BAO data set we find a peak at $\beta>0$. This data set seems to prefer a decrease of the DM mass with the expansion. The constraints on $\Omega_{\rm ede}^{\rm RD}$ are similar to those obtained in the absence of coupling.

\begin{table}[t!]
\centering
\begin{tabular}{|c||c|c|}
 \multicolumn{1}{c}{} & \multicolumn{1}{c}{} & \multicolumn{1}{c}{} \\%
\multicolumn{1}{c}{} &  \multicolumn{2}{c}{Planck18+SNIa+BAO}
\\\hline
{\small Parameter} & {\small ULA} & {\small CDE\_ULA}
\\\hline
$10^2\omega_b$ & $2.265^{+0.016}_{-0.022}$ (2.269) & $2.271^{+0.016}_{-0.022}$ (2.290)  \\\hline
$\omega_{\rm cdm}$ & $0.1225^{+0.0016}_{-0.0041}$ (0.1237)  & $0.1211^{+0.0011}_{-0.0029}$ (0.1201) \\\hline
$n_s$ & $0.974^{+0.005}_{-0.009}$ (0.978) & $0.974^{+0.005}_{-0.007}$ (0.978) \\\hline
$\tau_{\rm reio}$ & $0.059\pm 0.007$ (0.060) & $0.054^{+0.009}_{-0.010}$ (0.049) \\\hline
$\sigma_{12}$ & $0.799\pm 0.007$ (0.796) & $0.822^{+0.018}_{-0.021}$ (0.839)  \\\hline
$H_0$ &  $69.08^{+0.56}_{-1.20}$ (69.75)  & $68.99^{+0.46}_{-0.81}$ (69.24)  \\\hline
$f_{\rm ede}$ [\%] & $<4.80<9.46$ (5.17) & $<2.24<7.70$ (1.13) \\\hline
$z_{max}$ & $7806^{+1000}_{-6000}$ (3874) & $5317^{+1600}_{-2500}$ (4724) \\\hline
$\theta_{ini}$ & $>2.01$ (2.63) & $>1.39$ (1.83)  \\\hline
$\beta$ & $-$ & $-0.017^{+0.110}_{-0.083}$ (-0.089) \\\hline\hline
${\rm log_{10}(}m{\rm/eV)}$ & $-26.66^{+0.34}_{-0.79}$ (-27.30) & $-26.86^{+0.34}_{-0.60}$ (-27.10) \\\hline
${\rm log_{10}(}f{\rm/eV)}$ & $26.51^{+0.21}_{-0.39}$ (26.51) & $26.52^{+0.42}_{-0.51}$ (26.63) \\\hline
$r_d$ & $145.26^{+2.04}_{-0.78}$ (144.37) & $145.72^{+1.51}_{-0.40}$ (145.89)  \\\hline
$M$ & $-19.376^{+0.017}_{-0.037}$ (-19.355) & $-19.380^{+0.012}_{-0.025}$ (-19.373) \\\hline
$S_{8}$ & $0.825^{+0.011}_{-0.012}$ (0.823) & $0.845^{+0.019}_{-0.021}$ (0.859)  \\\hline
$\sigma_8$ & $0.819^{+0.008}_{-0.012}$ (0.822) & $0.842^{+0.018}_{-0.021}$ (0.862)  
 \\\hline
$S_{12}$ & $0.811^{+0.010}_{-0.013}$ (0.811)  & $0.831^{+0.018}_{-0.021}$ (0.846)  \\\hline
$\Delta\chi^2_{\rm min}$ & 5.41  & 7.87\\\hline
$\Delta{\rm AIC}$ & -0.59  & -0.13
\\\hline
\end{tabular}
\caption{As in Tables \ref{tab:tableCDE} and \ref{tab:tableEXP}, but for ULA and CDE\_ULA.}
\label{tab:tableULA}
\end{table}

ULA offers the interesting possibility of generating a peak in the EDE fraction around $z_{\rm eq}$ while respecting the very tight constraints on $\Omega_{\rm ede}$ at recombination  \cite{Gomez-Valent:2021cbe} (see Fig. \ref{fig:fig6}).  We show in Table \ref{tab:tableULA} and the contours plots of Fig. \ref{fig:fig10} the constraints on the main parameters of the model obtained with the data set Planck18+SNIa+BAO. There is a large positive correlation between $f_{\rm ede}$ and the Hubble parameter. Values of $H_0\sim 71$ km/s/Mpc accompanied by values of $f_{\rm ede}\sim 10\%$, fall inside the $2\sigma$ region according to the marginalized posterior. This seems to render the $H_0$ tension at the $\sim 2\sigma$ level. However, it has been recently demonstrated in \cite{Herold:2021ksg,Gomez-Valent:2022hkb} that volume effects play an important role in the marginalization process in this model. The constraints obtained with the profile distributions, which are not subject to these issues, are quite different, and favor larger values of $f_{\rm ede}$ and $H_0$ (cf. Fig. 5 and Table II in \cite{Gomez-Valent:2022hkb}). The constraints derived with the marginalized posteriors read $f_{\rm ede}<0.048$ and $H_0=(68.4^{+1.3}_{-0.5})$ km/s/Mpc\footnote{Here we provide the location of the peak of the posterior for $H_0$ and the corresponding uncertainties, instead of the mean.}, whereas with the profile distribution we obtain $f_{\rm ede}=0.052^{+0.022}_{-0.021}$ and $H_0=(69.8^{+0.9}_{-1.0})$ km/s/Mpc. A peak in $f_{\rm ede}$ is found with the profile distribution approach, $2\sigma$ away from 0 \cite{Herold:2021ksg,Gomez-Valent:2022hkb}. 

\begin{figure}[t!]
\begin{center}
\includegraphics[width=3.1in,height=2.7in]{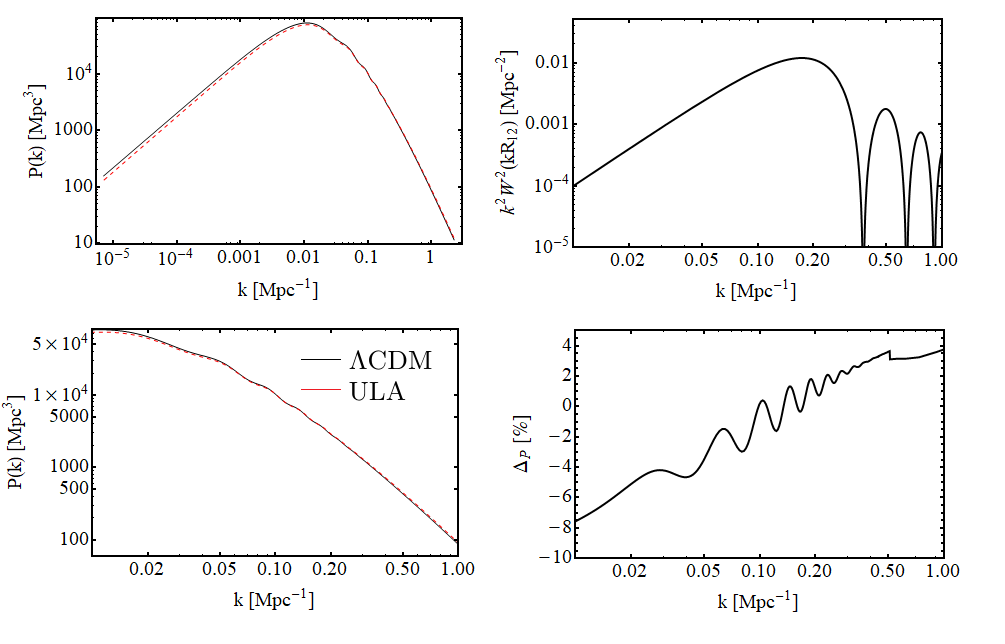}
\caption{{\it Left plots:} Comparison of the linear matter power spectrum of the $\Lambda$CDM and ULA using the best-fit values obtained in the fitting analysis with the Planck18+SNIa+BAO data set (cf. Sec. \ref{sec:data}); {\it Upper right plot:} This is the term with the window function that enters the equation of $\sigma_{12}$ \eqref{eq:rms}. Plotting this function is useful to see the range of $k$'s that is more important in the computation of the {\it rms} of mass fluctuations at $R_{12}=12$ Mpc; {\it Lower right plot:} Relative difference between the power spectra of the two models, $\Delta_P\equiv(P_{\rm ULA}-P_{\Lambda {\rm CDM}})/P_{\Lambda {\rm CDM}}$.}\label{fig:fig11}
\end{center}
\end{figure}

Regarding the LSS estimators, it is true that the model prefers values of $S_8$ and $\sigma_8$ larger than in the $\Lambda$CDM, but due to the points firstly raised in \cite{Sanchez:2020vvb} and subsequently  discussed also in \cite{Gomez-Valent:2022hkb} and Sec. \ref{sec:phenome} of this paper, we think it is safer to discuss the impact of the model on the linear perturbations in terms of the parameter $\sigma_{12}$, although we also provide the fitting values of $\sigma_8$, $S_8$ and $S_{12}$ in our tables. The constraint on $\sigma_{12}$ obtained in ULA is almost identical to the one obtained in the standard model, which means that ULA, when analyzed under the Planck18+SNIa+BAO data set, does not prefer an enhancement of the {\it rms} of mass fluctuations at scales of 12 Mpc. Actually one can see in the contours in the $(H_0,\sigma_{12})$-plane of Fig. \ref{fig:fig10} that it is possible to reach the $2\sigma$ band of SH0ES keeping a low $\sigma_{12}\sim 0.79-0.8$. In Fig. \ref{fig:fig11} we show the shape of the matter power spectrum obtained with the best-fit parameters in the $\Lambda$CDM and ULA. There is a significant relative decrease of power at large scales in ULA ($k\lesssim 0.1$ Mpc$^{-1}$), and an increase at low scales. This two opposite behaviors compensate each other in the computation of $\sigma_{12}$ (cf. formula \eqref{eq:rms}), yielding $\sigma_{12}\sim 0.8$. Although the model requires a larger value of $\omega_{\rm cdm}$ to counteract the early iSW effect caused by $f_{\rm ede}$, it is also able to accommodate a larger value of the cosmological constant, which is positively correlated with the current matter energy density. Despite being larger than in the $\Lambda$CDM, $\Omega^{(0)}_{m}$ still takes reasonable values, around $\sim 0.3$, see Fig. \ref{fig:fig12} ($\Omega_m^{(0)}\sim 0.26$ in the standard model). This larger matter fraction and the increase of $n_s$ cause the enhancement of $P(k)$ at large $k$'s, which is compensated by the decrease at low $k$'s that is induced by the bluer tilt of the primordial power spectrum. Constraints on the power spectrum at low scales will be important to further assess the viability of this model. According to the value of $\Delta {\rm AIC}$ (see again Table \ref{tab:tableULA}) the model is not statistically preferred under the Planck18+SNIa+BAO data set with respect to the $\Lambda$CDM, although it allows an alleviation of the $H_0$ tension.

If we leave the sum of the neutrino masses $\sum m_\nu$ free in the Monte Carlo analysis with ULA, we find very similar results to the case in which we use only one massive neutrino of $0.06$ eV (cf. Table \ref{tab:tableMassive}). We obtain $\sum m_\nu<0.13$ eV at $95\%$ c.l., which is very close to the upper bound derived with the $\Lambda$CDM, $\sum m_\nu<0.12$ eV. A larger mass of the neutrino masses increases the uncertainties of $\sigma_{12}$, but only slightly. This does not have a significant impact on the cosmological tensions, as it has been also recently reported in \cite{Reeves:2022aoi}, where the authors obtained $\sum m_\nu<0.15$ eV at $95\%$ c.l. using {\it Planck} and BOSS galaxy clustering data, and showed that the constraints on $\sum m_\nu$ are not affected by marginalization issues. We obtain a somewhat tighter constraint, closer to the lower bound of the inverted mass hierarchy.

\begin{figure}[t!]
\begin{center}
\includegraphics[width=3.3in,height=1.8in]{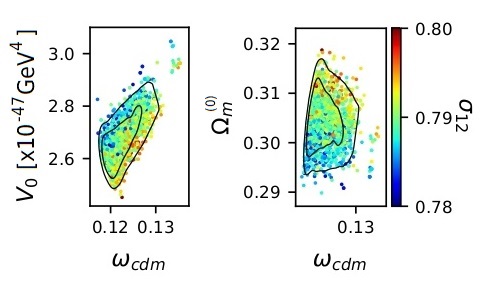}
\caption{Scatter and contour plots at $1\sigma$ and $2\sigma$ c.l. in the $(\omega_{\rm cdm},V_0)$ and $(\omega_{\rm cdm},\Omega^{(0)}_m)$ planes of the ULA model, obtained from the fitting analysis with the Planck18+SNIa+BAO dataset. We indicate the value of $\sigma_{12}$ at each point of the scatter plot. There is a significant positive correlation between the current CDM energy density and the value of the cosmological constant, which allows to have $\Omega_m^{(0)}\sim 0.3$ in the $1\sigma$ region even for large values of $\omega_{\rm cdm}\sim 0.13$. See the comments in the main text.}\label{fig:fig12}
\end{center}
\end{figure}

\begin{figure*}[t!]
\begin{center}
\includegraphics[width=6.1in,height=4.5in]{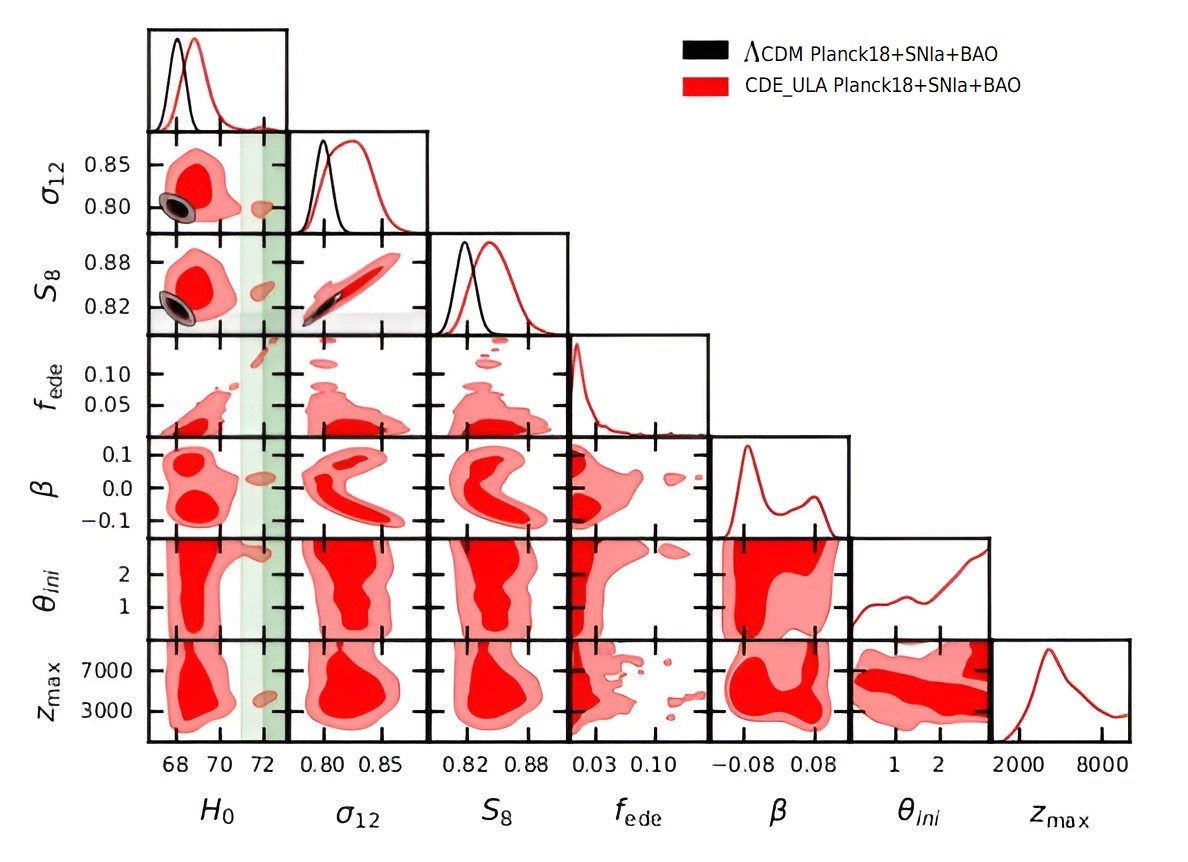}
\caption{Confidence regions at $1\sigma$ and $2\sigma$ c.l. and one-dimensional posterior distributions obtained with the Planck18+SNIa+BAO data set with the $\Lambda$CDM and CDE\_ULA.  }\label{fig:fig13}
\end{center}
\end{figure*}

If we extend the model to incorporate a coupling in the dark sector we do not gain that much. The value of $\Delta{\rm AIC}$ is very close to zero. The results for CDE\_ULA are shown in Table \ref{tab:tableULA} and Fig. \ref{fig:fig13}. As expected, when $\theta_{\rm ini}\lesssim \pi$ there is a symmetry between positive and negative values of the coupling, which is broken when $\theta_{\rm ini}$ is not sufficiently close to $\pi$. In all cases there is a preference for a decay of the dark energy mass with the cosmic expansion (as found also with CDE\_EXP), and this is why for $\theta_{\rm ini}\ll \pi$ the fit prefers a negative $\beta$. This explains why the peak at $\beta<0$ is larger. The latter is, though, still compatible with 0 at only $2\sigma$ c.l.

\begin{table*}[t!]
\centering
\begin{tabular}{|c ||c | c | c|}
 \multicolumn{1}{c}{} & \multicolumn{1}{c}{} & \multicolumn{1}{c}{} & \multicolumn{1}{c}{} \\ \multicolumn{1}{c}{} & \multicolumn{3}{c}{Planck18+SNIa+BAO [3 massive neutrinos]}
\\\hline
{\small Parameter} & {\small $\Lambda$CDM}  & {\small CDE\_const}& {ULA}
\\\hline
$10^2\omega_b$ & $2.248^{+0.014}_{-0.013}$ (2.246) & $2.238\pm 0.015$ (2.243) & $2.267^{+0.017}_{-0.022}$ (2.271)   \\\hline
$\omega_{\rm cdm}$ & $0.1187\pm 0.0008$ (0.1186) &  $0.1181^{+0.0010}_{-0.0008}$ (0.1177) & $0.1222^{+0.0016}_{-0.0040}$ (0.1257)   \\\hline
$n_s$ & $0.968\pm 0.004$ (0.970) & $0.966\pm 0.004$ (0.966) & $0.974^{+0.005}_{-0.010}$ (0.982)   \\\hline
$\tau_{\rm reio}$ & $0.059^{+0.007}_{-0.008}$ (0.064) & $0.058^{+0.007}_{-0.009}$ (0.056)  & $0.060^{+0.008}_{-0.009}$ (0.066)  \\\hline
$\sigma_{12}$ & $0.798^{+0.007}_{-0.008}$ (0.803) & $0.805^{+0.009}_{-0.011}$ (0.801) & $0.797^{+0.009}_{-0.008}$ (0.801)   \\\hline
$H_0$ & $67.89\pm 0.36$ (67.98) & $68.34^{+0.45}_{-0.55}$ (68.64) & $68.91^{+0.62}_{-1.20}$ (69.99)  \\\hline
$\sum m_\nu$ & $0.06<x<0.12$ (0.07) & $0.06<x<0.18$ (0.06)  & $0.06<x<0.13$ (0.07)   \\\hline
$f_{\rm ede}$ [\%] & $-$ & $-$  & $<5.05<9.47$ (7.59)  \\\hline
$z_{max}$ & $-$  & $-$  & $8200^{+1700}_{-6000}$ (6709)  \\\hline
$\theta_{ini}$ & $-$ & $-$ & $>2.11 (2.89)$  \\\hline
$\beta$ & $-$ & $0.033\pm 0.015$ (0.031) & $-$   \\\hline\hline
${\rm log_{10}(}m{\rm/eV)}$ & $-$ & $-$ & $-26.64^{+0.39}_{-0.82}$ (-26.78)  \\\hline
${\rm log_{10}(}f{\rm/eV)}$ & $-$ & $-$ & $26.49^{+0.27}_{-0.40}$ (26.43)  \\\hline
$r_d$ & $147.32^{+0.22}_{-0.21}$ (147.37) & $147.10^{+0.32}_{-0.23}$ (147.29) & $145.31^{+2.10}_{-0.80}$ (143.50)   \\\hline
$M$ & $-19.412^{+0.011}_{-0.010}$ (-19.410) & $-19.400^{+0.012}_{-0.015}$ (-19.393) & $-19.381^{+0.019}_{-0.037}$ (-19.346)   \\\hline
$S_{8}$ & $0.816^{+0.009}_{-0.010}$ (0.821) & $0.820\pm 0.011$ (0.813) & $0.822^{+0.012}_{-0.013}$ (0.833)   \\\hline
$\sigma_8$ & $0.808\pm 0.007$ (0.814) & $0.819^{+0.010}_{-0.013}$ (0.817) & $0.815^{+0.010}_{-0.013}$ (0.829)   
 \\\hline
$S_{12}$ & $0.800\pm 0.008$ (0.806) & $0.806^{+0.010}_{-0.011}$ (0.801) & $0.808^{+0.011}_{-0.014}$ (0.820)   \\\hline
$\Delta\chi^2_{\rm min}$ & $-$ & 1.25  & 3.94
\\\hline
$\Delta {\rm AIC}$ & $-$ & -2.75  & -4.06
\\\hline
\end{tabular}
\caption{As in the previous tables, but for the three models: $\Lambda$CDM, CDE\_const and ULA, with the Planck18+SNIa+BAO data set, and allowing the sum of the neutrino masses $\sum m_\nu$ to vary in the Monte Carlo, as explained in Sec. \ref{sec:data}. The constraints on this parameter are given in eV at $95\%$ c.l. See the discussion of these results in Sec. \ref{sec:results}.}\label{tab:tableMassive}  
\end{table*}

\begin{figure*}[t!]
\begin{center}
\includegraphics[width=7.3in,height=5in]{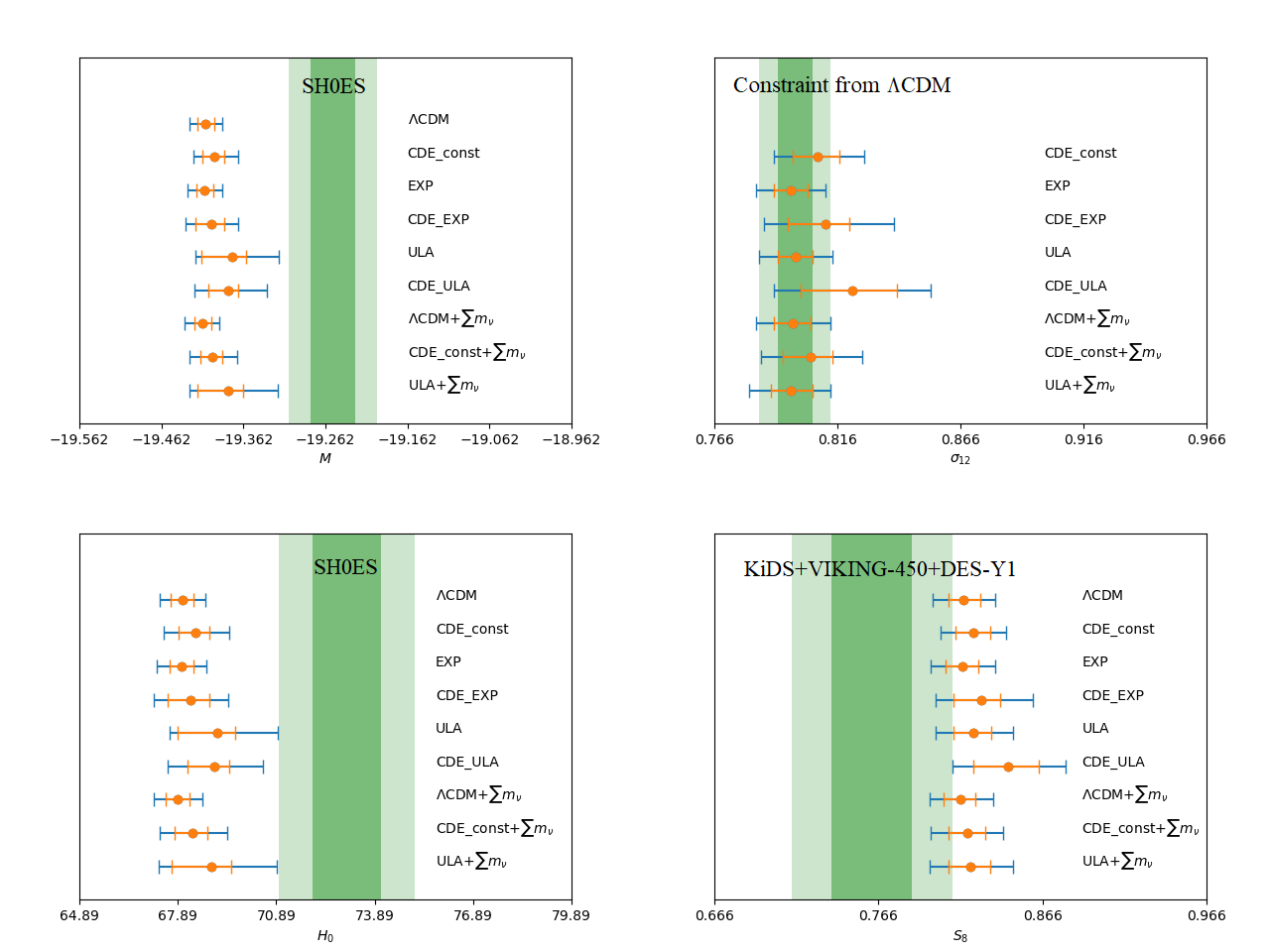}
\caption{{\it Left plots:} Marginalized constraints on $M$ and $H_0$ (in km/s/Mpc) at 1$\sigma$ and 2$\sigma$ c.l. for all the models studied in this paper (cf. Sec. \ref{sec:phenome}), obtained with the Planck18+SNIa+BAO data set. The green bands correspond to the baseline values of $M$ and $H_0$  measured by SH0ES \cite{Riess:2021jrx}. ULA is the model that alleviates the Hubble tension the most. No significant improvement is achieved by considering a coupling in the dark sector (CDE\_ULA) or more massive neutrinos (ULA+ $\sum m_{\nu}$); {\it Upper right plot:} Same as the left plots, but for $\sigma_{12}$. The bands in this case are taken from the marginalized Planck18+SNIa+BAO constraint on $\sigma_{12}$ in $\Lambda$CDM, see Table \ref{tab:tableCDE}. The constraint on $\sigma_{12}$ from ULA is almost identical to the one from the standard model; {\it Lower right plot:} Marginalized constraint on $S_8$ at 1$\sigma$ and 2$\sigma$ c.l. for all the models. The green band indicates the combined measurement of $S_8$ by KiDS+VIKING-450 + DES-Y1, again under the assumption of the standard model \cite{Joudaki:2019pmv}.} \label{fig:fig15}
\end{center}
\end{figure*}

It is clear from the contours in the $(\beta,\sigma_{12})$-plane that when we move away from the uncoupled scenario, i.e. from $\beta=0$, $\sigma_{12}$ grows. Large values of $\sigma_{12}\sim 0.85$ are not excluded by the Planck18+SNIa+BAO data set. As already mentioned, the latter prefers values of $\beta$ $\sim 2\sigma$ away from 0, and this is what generates a broader posterior for $\sigma_{12}$, peaking also at larger values compared to the uncoupled ULA model. The required values of $f_{\rm ede}$ to alleviate the Hubble tension can only be obtained in the vicinity of $\beta=0$, so the preference of the data for a non-null coupling, despite being still mild, inevitably leads to a tighter constraint on the maximum EDE fraction, which is now below $7.7\%$ at $95\%$ c.l. Unfortunately, in the absence of a prior from SH0ES, we do not find an alleviation of the tensions caused by the coupling. We have verified that these results are not strongly affected by volume effects.  

Before closing the discussion on our results, we would like to remark that the quantification of the Hubble tension, when formulated as a tension between the value of the absolute value of SNIa $M$ obtained from the distance ladder by SH0ES and the value obtained from the fitting analysis involving high-$z$ supernovae, give similar but not exactly equal results \cite{Gomez-Valent:2021cbe}. The latter tends to lead to slightly larger estimates of the tension. In the left plots of Fig. \ref{fig:fig15} we summarize our constraints on $H_0$ and $M$ for all the models studied in this paper, and compare them to the measurement carried out by the SH0ES Team \cite{Riess:2021jrx}. We find that ULA is the model that alleviates more the $H_0$ tension, and that neither a constant coupling between the dark components nor a larger sum of the neutrino masses can help to further mitigate it.

In the right plots of Fig. \ref{fig:fig15} we summarize the constraints on $\sigma_{12}$ and $S_8$ obtained for the various models. Some comments are in order. All the values of these LSS estimators are compatible with the result in the $\Lambda$CDM. There is no model leading to a significant decrease of these quantities. The constraints on $\sigma_{12}$ in ULA and ULA+$\sum m_\nu$ are actually almost identical to those found in the standard model. The estimation of the level of tension between the models and the weak lensing measurement from KiDS+VIKING-450+DES-Y1 is not straightforward and can be misleading for the issues discussed in \cite{Sanchez:2020vvb}, and also in Sec. \ref{sec:phenome}. Moreover, it is important to bear in mind that the observational constraint on $S_8$ from KiDS+VIKING-450+DES-Y1 (see the green bands shown in the lower right plot of Fig. \ref{fig:fig15}) has been obtained under the assumption of the $\Lambda$CDM, so it could lead to an overestimation of the level of tension in the context of non-standard cosmologies.


\section{Conclusions}\label{sec:conclusions}

In this paper we have studied some coupled and uncoupled early dark energy (EDE) models with different shapes of the EDE fraction, confronting them with the CMB data from {\it Planck}, the Pantheon compilation of supernovae of Type Ia and data on baryon acoustic oscillations, without including any prior on $H_0$ or $M$ from SH0ES. We have considered three different forms of the potential energy density for the scalar field, to wit: a constant potential, an exponential potential \eqref{eq:expPot} that produces a scaling regime in the matter- and radiation-dominated epochs, and an ultra-light axion-like (ULA) potential \eqref{eq:ULApotential} able to generate a peak in $\Omega_{\rm ede}$, typically around the matter-radiation equality time. We have discussed in detail the phenomenology of these models. Motivated by the fact that EDE has some difficulties in relieving simultaneously the $H_0$ and LSS tensions, we have explored in this paper the impact of a coupling between EDE and dark matter, and the presence of massive neutrinos. The coupling can make the dark matter mass to decay with the expansion, leading to a faster decrease of the dark matter energy density. This fact could in principle slow down the clustering of matter in the universe, and massive neutrinos could also help to suppress the amount of large-scale structure at low scales. Nevertheless, the coupling can also enhance the matter power spectrum due to the fifth force. In order to elucidate their net effect, we have put these ideas to the test, by performing dedicated fitting analyses. 

When the EDE potential is constant we find under the Planck18+SNIa+BAO data set a preference for a non-null coupling at $2\sigma$ c.l. This result resonates well with previous studies in the literature \cite{Pettorino:2013oxa,Planck:2015bue,Gomez-Valent:2020mqn}, and is not caused by volume effects \cite{Gomez-Valent:2022hkb}. Nevertheless, if we penalize the additional complexity of the model through the calculation of the AIC we find that the improvement in the description of the data is not sufficient to justify the need of the interaction in the dark sector. The $H_0$ tension is alleviated, yes, but persists at $\sim 3\sigma$. Massive neutrinos give some room for lower values of $\sigma_{12}$, but the gain is not substantial, and is accompanied by a slight decrease of $H_0$ as well. The upper bound of the sum of the neutrino masses reads $\sum m_\nu<0.18$ eV at $2\sigma$ c.l., and is considerably larger than in the $\Lambda$CDM, $\sum m_\nu<0.12$ eV. This might indicate that probing the neutrino mass hierarchy with cosmological data in a model-independent way might be quite difficult. Constraints obtained in the context of other models can be even more loosen, see e.g. \cite{SolaPeracaula:2020vpg}. This complication will presumably remain with the advent of future data. 

Moreover our results show that EDE models with an exponential potential have almost no impact on the cosmological tensions. We already reached this conclusion in \cite{Gomez-Valent:2021cbe}, but here we have also checked the effects introduced by the coupling. The latter allows us to describe considerably better the CMB data and we even find a moderate $\sim 2\sigma$ hint for a non-null interaction in the dark sector when only the {\it Planck} likelihood is considered. Nevertheless, this signal is diluted when we combine CMB with the SNIa and BAO data sets, and the tensions remain large. 

ULA is able to decrease the $H_0$ tension below the $2\sigma$ level, basically because it allows to have a larger EDE fraction around the matter-radiation equation time while respecting the stringent bounds at recombination \cite{Gomez-Valent:2021cbe}. The tension is even lower when it is quantified with the profile distribution method \cite{Gomez-Valent:2022hkb}.  The values of the LSS estimator $\sigma_{12}$ are similar to those found in the standard model. We have seen, though, that the shape of the power spectrum exhibits some important differences at small and large scales, so it could be useful to prove in the future the non-linear regime to study the viability of this model. For instance, there could be significant differences between the number counts of clusters in ULA and the $\Lambda$CDM obtained using the best-fit values of the parameters from our analysis. By allowing the mass of the neutrinos to take larger values in the fitting analysis we get an upper bound which is extremely close to the one obtained in the $\Lambda$CDM, $\sum m_\nu<0.13$ eV at $2\sigma$ c.l., and its posterior peaks very close to the minimum value allowed by the experiments on atmospheric and solar neutrinos, at (0.07 eV). Massive neutrinos do not induce any significant shift on the other parameters and, hence, they are unable to lower the tensions. This has been reported recently also in \cite{Reeves:2022aoi}, and this conclusion is not affected by marginalization issues of any kind. Unfortunately, when we activate the coupling in the context of ULA we do not appreciate any improvement regarding the tensions either. 

The cosmological tensions tackled in this paper cannot be fully alleviated by an interaction with a constant coupling between the two dark components nor neutrinos with larger masses. In the best case, namely ULA, the $H_0$ tension is reduced to $\sim 2\sigma$, and the amount of clustering at linear scales is similar to the $\Lambda$CDM. A more detailed treatment of the non-linear scales in these models and the inclusion of weak lensing and galaxy clustering data in the analysis could shed more light on the status of the tensions, of course. This could be done by performing scale cuts on CMB and weak lensing data to explore only the effect of linear scales or by modeling accurately the non-linear power spectrum (on the lines of e.g. \cite{Casas:2015qpa}) to exploit the statistical information contained in the full range of scales covered by the CMB and LSS data. This goes beyond the scope of this paper, and is left for a future work.


\vspace{0.5cm}
\noindent {\bf Acknowledgements}
\newline
\newline
\noindent AGV is funded by the Istituto Nazionale di Fisica Nucleare (INFN) through the project of the InDark INFN Special Initiative: ``Dark Energy and Modified Gravity Models in the light of Low-Redshift Observations'' (n. 22425/2020). ZZ is supported by the DFG Research Training Group “Particle Physics beyond the Standard Model” (GRK 1940). LA acknowledges support from DFG project  456622116. The authors would like to thank Dr. Elmar Bittner for his precious technical help in the use of the cluster of the ITP Heidelberg, which has been crucial for the completion of this project.


\appendix 

\begin{figure}[t!]
\begin{center}
\includegraphics[width=3.5in, height=4in]{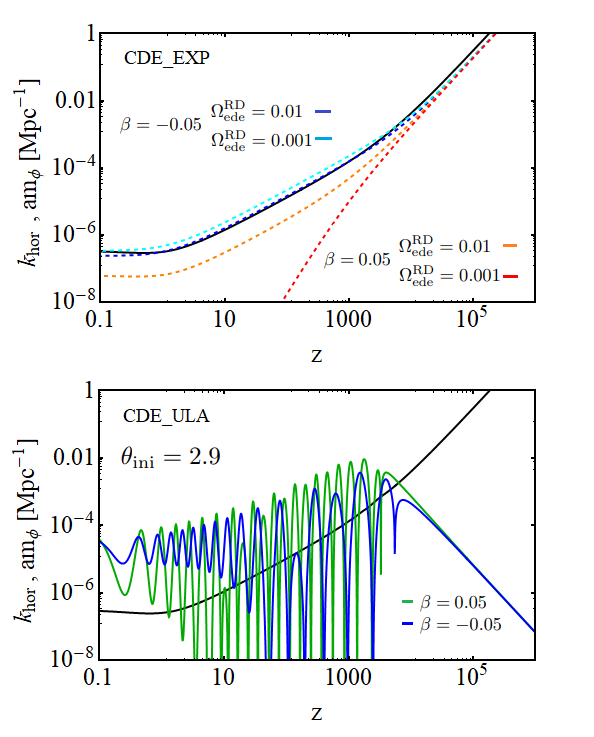}
\caption{Comoving wave mode at horizon crossing, $k_{\rm hor}$, and comoving dark energy mass, $am_\phi$, as a function of the redshift for several CDE\_EXP (upper plot) and CDE\_ULA (lower plot) models. The parameters that are not specified explicitly in the legends are set as in Fig. \ref{fig:fig2} and the left plots of Fig. \ref{fig:fig6}, respectively.}\label{fig:figAppendixA}
\end{center}
\end{figure}

\section*{Appendix A: Impact of the dark energy mass on the LSS}

In this brief appendix we discuss the impact that the non-zero dark energy mass $m^2_\phi=d^2V/d\phi^2$ has on the LSS both in CDE\_EXP and CDE\_ULA. In the upper plot of Fig. \ref{fig:figAppendixA} we can see that for typical values of $\Omega_{\rm ede}^{\rm RD}$ and the coupling in the CDE\_EXP models, the comoving modes that are relevant for the LSS, i.e. $10^{-2}\,{\rm Mpc}^{-1}\lesssim k\lesssim 1\,{\rm Mpc}^{-1}$, are already larger than the comoving mass of the DE at the horizon crossing, i.e. when the modes of interest are equal to $k_{\rm hor}$ in the plot. The corresponding comoving scales rapidly become much smaller than the range of the fifth force. This means that the mass terms $m_\phi$ in the equation of the density contrast of DM \eqref{eq:densityContrastDM} can be safely neglected in this model.

\begin{table}[t!]
\centering
\begin{tabular}{|c || c| c| c| c|   }
\multicolumn{1}{c}{} & \multicolumn{4}{c}{Planck18} 
\\\hline
{\small $\Delta\chi^2_i$}  & {$\Lambda$CDM} & {\small CDE\_const} & {\small EXP} & {\small CDE\_EXP} 
\\\hline
{\small CMB highl} & 2351.83  & 2.00 & 3.19 & 7.54  
\\\hline
{\small CMB EE lowl} & 396.03 & 0.32  & -0.13 & -0.59 
\\\hline
{\small CMB TT lowl} & 23.27  & -0.08 & -0.21 & 0.72 
\\\hline
{\small CMB lens} & 8.81 & -0.31 & -0.10 & -0.16 
\\\hline
{\small Total} & 2779.94  & 1.93 & 2.75 & 7.51 
\\\hline
\end{tabular}
\caption{Differences $\Delta\chi^2_{i,j}\equiv \chi^2_{i,\Lambda}-\chi^2_{i,j}$ for the individual data sets $i$ and non-standard cosmological models $j$, using the best-fit values obtained in the analyses with Planck18 data.  In the last row we show the total differences, $\Delta\chi^2_{{\rm min},j}=\chi^2_{\Lambda,{\rm min}}-\chi^2_{j,{\rm min}}$. For the $\Lambda$CDM, instead, we show in the first column the individual values of $\chi^2_{i,\Lambda}$, and their sum.}
\label{tab:tableB1}
\end{table}

In the CDE\_ULA model the evolution of the DE mass is very different from the one in CDE\_EXP. The aforementioned modes do not feel the non-null DE mass when these scales reenter the horizon, since for these modes $k\gg am_\phi$ already at the crossing time. This is shown in the lower plot of Fig. \ref{fig:figAppendixA}. The scales that reenter the horizon at $30\lesssim z\lesssim 10^{3}$, though, are sensitive to the oscillatory behavior of the DE mass. Nevertheless, these modes are not relevant for the LSS, and the mean value of $am_\phi$ is still much smaller than $k_{\rm hor}$.      


\section*{Appendix B: Shooting method for ULA and CDE\_ULA}
We implement a shooting method in ULA and CDE\_ULA in order to map every combination of the input parameters $(f_{\rm ede},z_{\rm max})$ to the corresponding parameters $(m,f)$ of the ULA potential, see formula \eqref{eq:ULApotential} in Sec. \ref{sec:phenome}. We apply the following steps:

\begin{table*}[t!]
\centering
\begin{tabular}{|c ||c | c | c| c| c| c|  }
\multicolumn{1}{c}{} & \multicolumn{6}{c}{Planck18+SNIa+BAO} 
\\\hline
{\small $\Delta\chi^2_i$} & {$\Lambda$CDM} & {\small CDE\_const} & {\small EXP} & {CDE\_EXP} & {\small ULA} & {\small CDE\_ULA} 
\\\hline
{\small CMB highl} & 2349.83 & 1.19 & -0.4 & 2.28 & 2.58 & 4.33  
\\\hline
{\small CMB EE lowl} & 397.90 &  1.87 & 1.12 & 2.19 & 0.84 & 2.16
\\\hline
{\small CMB TT lowl} & 23.51 & -0.79  & 0.73 & 1.06 & 1.88 & 2.08
\\\hline
{\small CMB lens} & 9.14 & 0.20 & 0.22 & 0.05 &  -0.39 & -1.29
\\\hline
{\small SNIa} & 1025.77 & -0.26  & -1.19 & -0.16 & 0.03 & 0.14
\\\hline
{\small BAO} & 9.45 & 0.19 & -0.65 & -1.71 & 0.43 & 0.42
\\\hline
{\small Total} & 3815.61 & 2.39  & -0.15 & 3.75 & 5.41 & 7.87
\\\hline
\end{tabular}
\caption{As in TABLE \ref{tab:tableB1}, but for the analyses with the Planck18+SNIa+BAO data set.}
\label{tab:tableB2}
\end{table*}

\begin{table}[t!]
\centering
\begin{tabular}{|c ||c | c | c|  } \multicolumn{4}{c}{Planck18+SNIa+BAO [3 massive neutrinos]} 
\\\hline
{\small $\Delta\chi^2_i$} & {$\Lambda$CDM} &  {\small CDE\_const} & {\small ULA} 
\\\hline
{\small CMB highl} & 2348.37 & -1.72 & 3.88   
\\\hline
{\small CMB EE lowl} & 399.14 & 2.71 & -0.24  
\\\hline
{\small CMB TT lowl} & 22.76 & -0.68  & 1.44  
\\\hline
{\small CMB lens} & 8.77 & -0.74 & -0.41 
\\\hline
{\small SNIa} & 1025.75 & 0.12  & -0.13 
\\\hline
{\small BAO} & 9.64 & 0.50 & 0.42 
\\\hline
{\small  $\Delta m^2_{21},\Delta m^2_{31}$} & 0.23 & 0.10 & -1.18 
\\\hline
{\small Total} & 3814.66  & 1.25 & 3.94  
\\\hline
\end{tabular}
\label{tab:tableB3}            
\caption{As in TABLE \ref{tab:tableB2}, but allowing the neutrino masses to vary in the Monte Carlo runs. $\Delta m^2_{21},\Delta m^2_{31}$ refer to the priors employed for these quantities, see Sec. \ref{sec:data}.}
\end{table}

\begin{itemize}

\item We start with an initial guess for $\Big({\rm log_{10}(}m{\rm/eV)},{\rm log_{10}(}f{\rm/eV)}\Big)$. We build a grid around it and find the location of $x^{(1)}\equiv\Big({\rm log_{10}(}m{\rm/eV)^{(1)}},{\rm log_{10}(}f{\rm/eV)^{(1)}}\Big)$ that leads to the minimum $\chi^2$ value on the grid. The associated values of the cosmological (input) parameters are labeled as $y^{(1)}\equiv\Big(f_{\rm ede}^{(1)}, \rm log_{10}(z_{max})^{(1)}\Big)$.  This step is important to start the main part of the shooting sufficiently close to our final target $y^{(F)}$.

\item We build the Jacobian matrix $J_{ij}^{(1)}\equiv\frac{\partial y_i^{(1)}}{\partial x_j^{(1)}}$, approximating it as follows,

\begin{equation}
J^{(1)}\approx
\begin{pmatrix}
\frac{y_1(x_1+\delta x_1,x_2)-y_1^{(1)}}{\delta x_1} & \frac{y_1(x_1,x_2+\delta x_2)-y_1^{(1)}}{\delta x_2}   \\
\frac{y_2(x_1+\delta x_1,x_2)-y_2^{(1)}}{\delta x_1} & \frac{y_2(x_1,x_2+\delta x_2)-y_2^{(1)}}{\delta x_2} \\
\end{pmatrix}\,,\nonumber
\end{equation}
with $\delta x_1$ and $\delta x_2$ two sufficiently small increments.

\item A new trial $x^{(2)}$ is related to $y^{(\rm F)}$ and the Jacobian matrix $J^{(1)}$ through the relation:

\begin{equation}\label{eq:shooting}
   x^{(2)}= x^{(1)}+\Big(J^{(1)}\Big)^{-1}(y^{\rm (F)}-y^{(1)}),
\end{equation}
where the matrix $\Big(J^{(1)}\Big)^{-1}$ is the inverse of $J^{(1)}$. Eq. \eqref{eq:shooting} is nothing else than the Taylor expansion of $x$ around $y^{(1)}$ evaluated at $y^{(F)}$, truncated at linear order. We compute the new Jacobian matrix $J^{(2)}$ following the previous steps to obtain $x^{(3)}$, and we repeat this iterative process until reaching convergence (within the desired precision). 
\end{itemize}
\vspace{0.3cm}
\section*{Appendix C: Tables with the individual $\Delta\chi^2_{i}$}

This appendix contains Tables \ref{tab:tableB1}-\ref{tab:tableB3}, where we list the contribution from the individual data sets to the total $\chi^2_{\rm min}$ for each model and fitting analysis carried out in this work. These tables complement the information reported in the tables of Sec. \ref{sec:results}.


\bibliographystyle{apsrev4-1}
\bibliography{coupledEDE_neutrinos}

\end{document}